\documentclass[preprint]{aastex}
\usepackage{emulateapj5}
\usepackage{apjfonts}
\usepackage{natbib}

\input psfig.tex

\def\aa{{A\&A}}
\def\aas{{ A\&AS}}
\def\aj{{AJ}}

\def\amin{$^\prime$}
\def\annrev{{ARA\&A}}
\def\apj{{ApJ}}
\def\apjs{{ApJS}}

\def\deg{$^{\circ}$}

\def\kms{km s$^{-1}$}

\def\lax{{$\mathrel{\hbox{\rlap{\hbox{\lower4pt\hbox{$\sim$}}}\hbox{$<$}}}$}}
\def\gax{{$\mathrel{\hbox{\rlap{\hbox{\lower4pt\hbox{$\sim$}}}\hbox{$>$}}}$}}
\def\simlt{\lower.5ex\hbox{$\; \buildrel < \over \sim \;$}}
\def\simgt{\lower.5ex\hbox{$\; \buildrel > \over \sim \;$}}

\def\mbh{{$M_{\rm BH}$}}
\def\micron{{$\mu$m}}
\def\mnras{{MNRAS}}
\def\nat{{Nature}}
\def\pasp{{PASP}}

\def\percm2{cm$^{-2}$}

\def\solmass{$M_\odot$}

\def\hi{\ion{H}{1}}

\def\mhi{$M_{{\rm H~I}}$}
\def\lb{$L_B$}

\def\vm{${\upsilon_m}$}

\def\sig{$\sigma_0$}

\slugcomment{To appear in {\it The Astrophysical Journal}.}
\lefthead{Ho}
\righthead{Bulge and Halo Kinematics}

\begin{document}

\title{Bulge and Halo Kinematics Across the Hubble Sequence}

\author{Luis C. Ho}

\affil{The Observatories of the Carnegie Institution of Washington, 813 Santa 
Barbara St., Pasadena, CA 91101}

\begin{abstract}
The correlation between the maximum rotational velocity of the disk (\vm) and 
the central stellar velocity dispersion of the bulge (\sig) offers insights 
into the relationship between the halo and the bulge.  We have assembled 
integrated \hi\ line widths and central stellar velocity dispersions to study 
the \vm--\sig\ relation for 792 galaxies spanning a broad range of Hubble 
types.  Contrary to earlier studies based on much smaller samples, we find 
that the \vm--\sig\ relation exhibits significant intrinsic scatter and that 
its zeropoint varies systematically with galaxy morphology, bulge-to-disk 
ratio, and light concentration, as expected from basic dynamical 
considerations.  Nucleated but bulgeless late-type spiral galaxies depart 
significantly from the \vm--\sig\ relation.  While these results render 
questionable any attempt to supplant the bulge with the halo as the 
fundamental determinant of the central black hole mass in galaxies, the 
observed distribution of \vm/\sig, which depends on both the density profile 
and kinematic structure of the galaxy, offers a useful constraint on galaxy 
formation models.  With the aid of a near-infrared Tully-Fisher relation, we 
identify a population of otherwise normal, luminous galaxies that have 
exceptionally low values of \vm/\sig.  We argue that a significant fraction of 
the \hi\ gas in these kinematically anomalous objects is dynamically 
unrelaxed, having been acquired externally either through capture from tidal 
interactions or through cold accretion from the intergalactic medium.
\end{abstract}


\section{Introduction}

It is accepted that every galaxy contains an extended dark matter halo, but 
precisely how the halo couples to the luminous components of the galaxy 
remains a subject of lively debate.  Much attention has been given to the 
connection between the disk and the halo, in particular as it concerns the 
fraction of baryons that collapse to form the disk and the role of adiabatic 
contraction (e.g., Navarro \& Steinmetz, 2000; Dutton et al. 2007).  In disk 
galaxies that contain bulges, what is the relationship between the bulge and 
the halo?  While a variety of methods can be used to address this question, 
important insights can be gained by investigating the relation between the 
line-of-sight central stellar velocity dispersion of the bulge, \sig, and the 
deprojected maximum rotation velocity of the disk, \vm, which effectively 
traces the circular velocity of the halo.  Whitmore and collaborators 
(Whitmore et al.  1979; Whitmore \& Kirshner 1981) first looked into this 
issue, using a small sample of spiral galaxies for which they had \sig\ and 
\vm\ measured through integrated \hi\ velocity profiles.  They find 
$\upsilon_m/\sigma_0 \approx 1.2$ to 2, with the ratio increasing with 
decreasing bulge-to-disk ratio; elliptical galaxies roughly follow the same 
pattern (Fall 1987; Franx 1993).  These trends were subsequently confirmed for 
a larger sample of normal (Whittle 1992a) and Seyfert (Nelson \& Whittle 1996) 
galaxies.

The \vm--\sig\ relation recently has attracted renewed attention in the 
context of black hole demographics studies.  The interest is three-fold.  
First, if \sig\ is related to \vm, then the existence of the 
$M_{\rm BH}-\sigma_0$ relation (Gebhardt et al. 2000; Ferrarese \& Merritt 
2000) suggests that the black hole mass may be more fundamentally tied to the 
halo mass rather than the bulge mass.  This is the argument made by Ferrarese 
(2002), who revisited the \vm--\sig\ correlation originally introduced by 
Whitmore and collaborators.  Second, apart from its theoretical implications, 
the existence of a \vm--\sig\ correlation presents a new empirical tool for 
black hole demographics studies.  Stellar velocity dispersions are not always 
easy or even possible to obtain, especially for active galaxies, whose bright 
nonstellar nuclei often overwhelm the stellar continuum and make measurement 
of central stellar velocity dispersions exceedingly challenging (e.g., Greene 
\& Ho 2006).  In such circumstances, it may be more feasible to measure \vm\ 
for the disk, either through spatially resolved rotation curves or 
integrated \hi\ line profiles (Ho et al. 2007a).  Finally, the \vm--\sig\ 
correlation represents a new scaling relation for galaxies, which, like other 
more familiar scaling relations, serves as an important boundary condition for 
theoretical models of galaxy formation.

Ferrarese (2002) compiled kinematic data for 16 disk (spiral and S0) galaxies 
with dynamical determinations of black hole masses to show that the rotation 
velocity of the disk, measured on the flat part of the rotation curve, follows 
a tight, nearly linear correlation with the central stellar velocity 
dispersion of the bulge.  For her sample, the correlation breaks down for 
$\sigma_0$ \lax\ 80 \kms.  The 19 elliptical galaxies with rotation velocities 
derived from the dynamical models of Kronawitter et al. (2000) and Gerhard et 
al. (2001) seem to fall on the same correlation, suggesting that the 
\vm--\sig\ relation is universally obeyed by galaxies of all types.  This 
result was echoed in subsequent studies (Baes et al. 2003), some of which 
(Pizzella et al. 2005; Buyle et al. 2006) additionally suggested that 
low-surface brightness galaxies follow a separate, nearly parallel relation 
compared to high-surface brightness galaxies by having a larger \vm\ for a 
given \sig.  In the most comprehensive analysis to date, Courteau et al. 
(2007) challenged the existence of a tight, Hubble type-invariant \vm--\sig\ 
relation.  Instead, their enlarged sample clearly demonstrates that \vm/\sig\ 
systematically varies with the concentration of the galaxy light profile, a 
result already apparent in the original work of Whitmore and others.  

This paper examines the \vm--\sig\ correlation using an extensive sample of 
nearby galaxies with accurate measurements of 

\begin{figure*}[t]
\centerline{\psfig{file=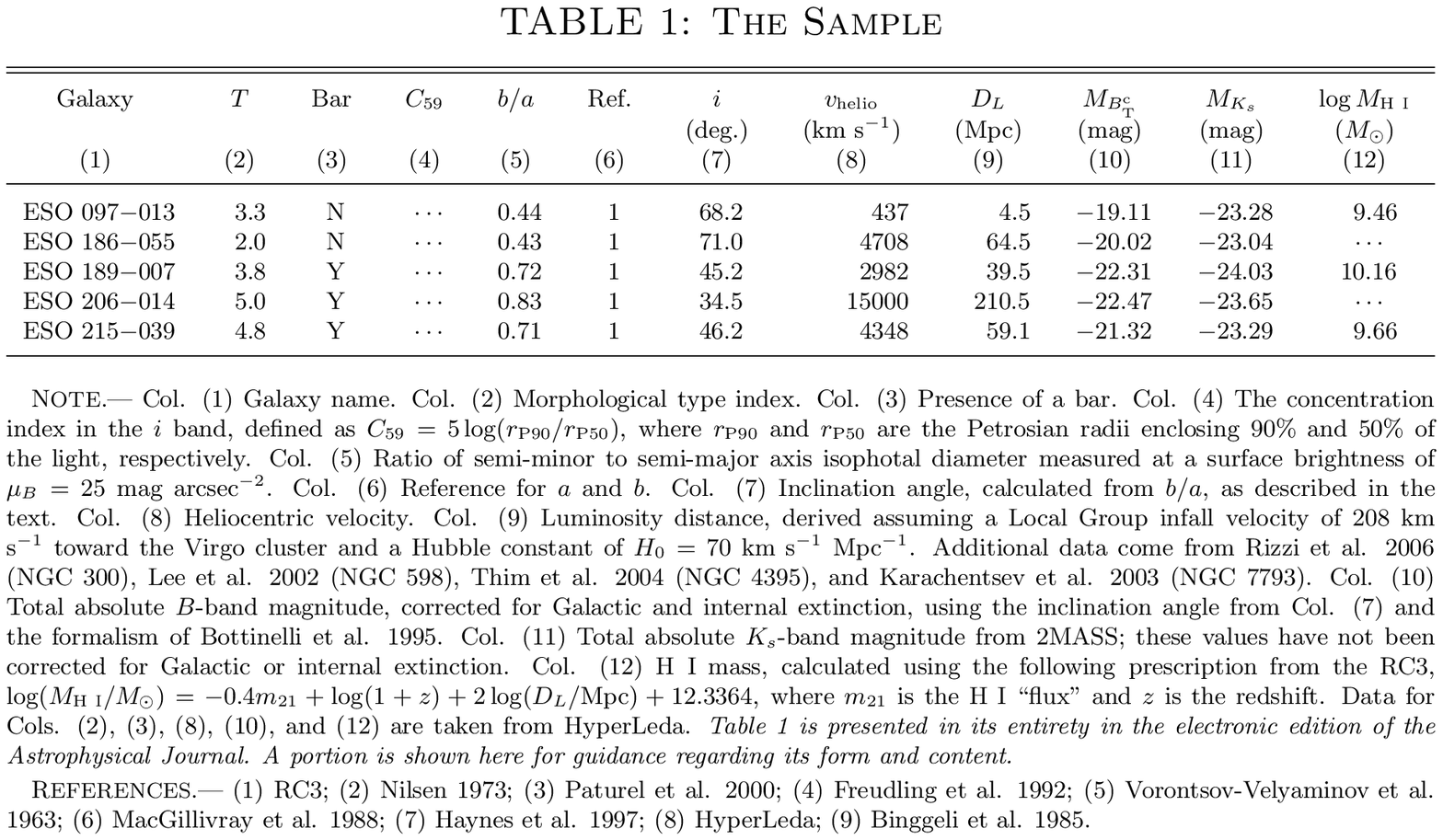,width=18.5cm,angle=0}}
\end{figure*}

\noindent 
both central stellar velocity 
dispersions and disk rotation velocities derived from integrated \hi\ line 
widths.  Our sample is significantly larger than those employed in recent 
studies, and it covers galaxies spanning a very wide range in Hubble types, 
allowing us to investigate trends with bulge-to-disk ratio.   This analysis is 
similar in spirit to that of Courteau et al. (2007), the main difference being 
that ours utilizes integrated \hi\ line widths instead of resolved rotation 
curves to estimate \vm, a shortcut that enables us to dramatically increase 
the sample size.   

We confirm, with greater statistical weight, that the \vm--\sig\ relation
systematically depends on the galaxy luminosity density profile, parameterized 
by Hubble type, bulge-to-disk ratio, or concentration index, and that there 
exists significant intrinsic scatter at a given density.  This finding does 
not bode well for attempts to use galaxy rotation velocities to predict black 
hole masses, nor does it support the contention that the black hole mass is 
more fundamentally linked to the halo than the bulge, but it does offer a new, 
potentially powerful constraint on galaxy formation models.  We draw attention 
to a subset of galaxies characterized by having exceptionally narrow \hi\ 
profiles for their central stellar velocity dispersion.  We suggest that the 
neutral hydrogen in these systems lies in a plane offset from that of the 
stellar disk, or that it has not yet settled into dynamical equilibrium with 
the stars.

\vskip 0.5cm
\section{Data Compilation}

Our goal is to compile as large a collection of galaxies as possible having 
relatively homogeneous, well-documented measurements of central stellar 
velocity dispersions and rotational velocities that probe the flat part of the 
disk rotation curve.  Apart from the sample size, a key difference between our 
analysis and those of previous studies [with the exception of initial work by 
Whitmore et al.  (1979), Whitmore \& Kirshner (1981), Whittle (1992a), and 
Nelson \& Whittle (1996)] is that our values of \vm\ come from spatially 
integrated (single-dish) \hi\ profiles.  Previous authors have stressed the 
importance of measuring \vm\ from spatially resolved, extended optical spectra 
that sample the flat part of the rotation curve.  Here we wish to emphasize 
that integrated \hi\ line widths provide a robust and efficient substitute for
estimating \vm.  This has been well-documented in numerous studies, many
motivated by the desire to use \hi\ line widths for distance-scale
investigations using the Tully-Fisher (Tully \& Fisher 1977) relation (e.g.,
Rubin et al. 1978; Thonnard 1983; Mathewson et al. 1992; Courteau 1997).  The
global \hi\ line width imprints both the shape of the galaxy's rotation curve
and the actual spatial distribution of the neutral hydrogen.  But since the
\hi\ distribution in spiral galaxies typically extends to twice the optical
radius (Broeils \& Rhee 1997; Noordermeer et al. 2005), in practice the width
of the \hi\ velocity profile is quite robust to different rotation curves and
\hi\ distributions (Roberts 1978).

We have compiled a database that consists of four samples, which we describe 
in turn.

\begin{itemize}

\item{{\it Sample 1 ---}\ \ \
The primary source of data for our investigation comes from Hyperleda\footnote{
{\tt http://leda.univ-lyon1.fr}} (Paturel et al. 2003a).  This catalog is 
periodically updated, and the entries used in the present analysis are 
reported to be current up to the end of 2003.  The stellar velocity 
dispersions in 

\begin{figure*}[t]
\centerline{\psfig{file=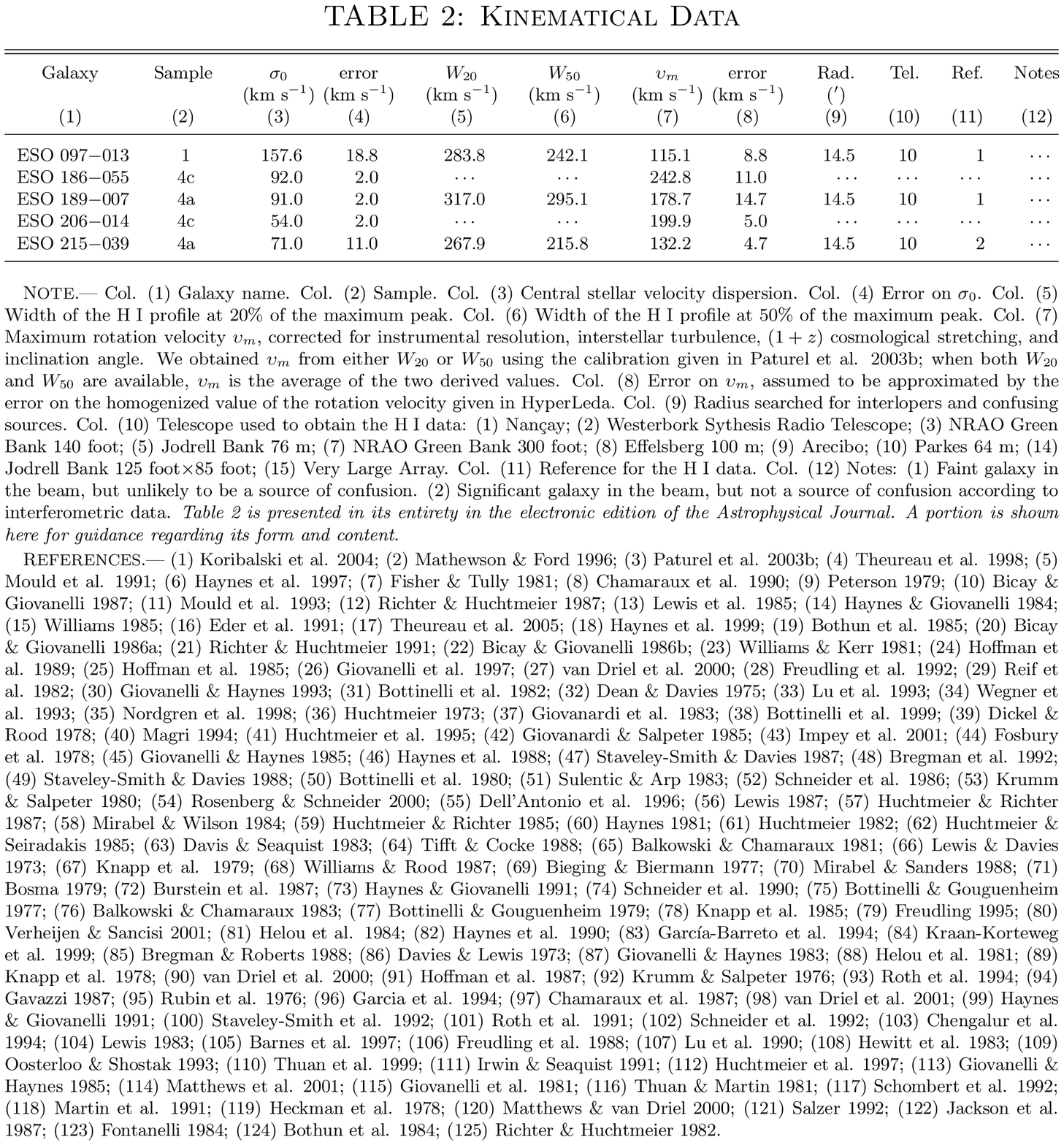,width=18.5cm,angle=0}}
\end{figure*}

\noindent
Hyperleda have been ``homogenized'' following the precepts of 
McElroy (1995).  Generally the measurements pertain to a central aperture that 
is smaller than the effective radius of the bulge.  Although it is sometimes 
customary to scale the velocity dispersions to a fixed aperture (e.g., to 
$R_{\rm eff}/8$; J{\o}rgensen et al. 1995), this practice assumes that all 
bulges possess a similar velocity dispersion profile, which, according to the 
observations of Pizzella et al. (2005), appears not to be the case.  We thus 
use the central values of $\sigma_0$, which for our sample on average have 
uncertainties of 13\%, with a standard deviation of 11\%.

Nearly all of the rotation velocities in Hyperleda are derived from
spatially integrated \hi\ line widths, homogenized in the manner described in
Paturel et al. (2003b). Using extensive sets of galaxies that contain both
integrated \hi\ profiles and spatially resolved optical rotation curves, these
authors determined the optimal transformation between the \hi\ line width
measured at different levels of the line peak (e.g., 20\% or 50\% of the
maximum) and the maximum velocity of rotation.  In this way, the archived \hi\
measurements, which can be quite voluminous for any given galaxy, can be
standardized and reduced to a single value for the rotation velocity. Although
this procedure is clearly very useful, following the spirit in which the
stellar velocity dispersions themselves were treated, unfortunately the
rotation velocities listed in Hyperleda cannot be used at face value because
of the possibility of source confusion.  Single-dish \hi\ measurements subtend
over a significant beam area, typically from 3\farcm5 for the Arecibo
telescope to as much as 22\amin\ for the Nan\c{c}ay telescope, within which
there is a nonnegligible probability of contamination from interlopers or
neighboring galaxies.  Hyperleda does {\it not}\ take this crucial effect into
consideration prior to homogenizing the \hi\ data.

Accordingly, we have taken the following steps to cull the data.  Beginning
with an initial sample of $\sim 1500$ galaxies having both velocity 
dispersions and \hi\ measurements, for each galaxy we systematically inspected 
the list of \hi\ line widths given in Hyperleda.  In this process, we give 
preference to more modern observations if available, to data taken with the 
highest spectral resolution, to line width measurements that pertain to either 
the 20\% or 50\% of the maximum of the line profile (preferably both), and, 
whenever possible, we try to minimize the heterogeneity of the sample by 
favoring larger, more systematic surveys.  For each galaxy, we identify what 
we deem to be the most robust line width measurement taken with the {\it 
smallest}\ available beam, where robustness is judged by whether the line 
width has reached a stable, asymptotic value when data from multiple 
telescopes are listed.  We then carefully examined digital optical images, 
from the SDSS if available or else from the scanned images of the Palomar 
Digital Sky Survey, in combination with redshift information listed in the 
NASA/IPAC Extragalactic Database (NED)\footnote{{\tt
http://nedwww.ipac.caltech.edu/}} to eliminate potential sources of confusion
within a radius equal to the full-width at half power of the beam of the
chosen telescope (see Table~2).  For observations taken with the Arecibo
telescope, the search radius was increased to 7\farcm5 to account for the
extended sidelobes of the beam; the intensity of the first sidelobes of the
beam drops to $\sim$10\% of the peak at a distance of 5\farcm5 from the beam
center, and by 7\amin--8\amin\ it becomes negligible (Heiles et al. 2000).  A
number of galaxies contain faint, low-surface brightness companions that
formally lie within the \hi\ beam, but generally such companions can be ruled
out as sources of confusion because of their low luminosities (they would
otherwise grossly violate the Tully-Fisher relation given the large line
widths).  Galaxies that are likely to be confused are flagged and omitted from
the sample.  We retained a few galaxies that, although potentially confused
within the single-dish beam, have line widths consistent with the velocity
amplitude measured in interferometer maps, which suggest that confusion is not
a problem.  To avoid complications in the interpretation of their \hi\
kinematics, we also removed galaxies known to be merger remnants or that, from
our visual inspection, otherwise possess disturbed morphologies, tidal tails,
shells, or polar rings.  We do not consider dwarf irregular galaxies 
($T \geq 9.0$) because they are unlikely to have trustworthy central stellar 
velocity dispersion measurements or unambiguous bulges.  
The above selection reduced the Hyperleda sample to 293 galaxies.
}

\item{{\it Sample 2 ---}\ \ \ Many galaxies in Hyperleda have usable \hi\ line 
widths but no stellar velocity dispersion in the database. A significant 
number of them overlap with the Sloan Digital Sky Survey (SDSS; York et al.  
2000), and stellar velocity dispersions for 435 of these have been measured 
and made publically available by D. J. 
Schlegel\footnote{{\tt http://spectro.princeton.edu}}.  In this selection, we 
only retain galaxies that have velocity dispersions larger than the spectral 
instrumental resolution of SDSS ($\sim 70$ \kms) and that satisfy our 
morphological type cut ($T < 9.0$).  We visually examined the SDSS spectrum of 
every object to confirm that the reported velocity dispersion is sensible;
a number of spurious values, almost all resulting from erroneous fitting
of noisy spectra of faint, dwarf galaxies, were rejected during this process.
D. J. Schlegel et al. (2007, in preparation; see also Heckman et al. 2004)
measure velocity dispersions using a direct-fitting algorithm and template
stars from a library of echelle stellar spectra.  The Appendix presents an
external comparison of the velocity dispersions given in SDSS with those
published in Hyperleda, for objects in common between samples 1 and 2.  The 
SDSS values are on average 12\% larger than those in Hyperleda.
}

\item{{\it Sample 3 ---}\ \ \ The issue of whether \sig\ or \vm\ more
fundamentally tracks \mbh\ can be most effectively addressed by using
extremely late-type spirals that essentially lack a bulge altogether (B\"oker
et al. 2003) and rarely contain a central black hole (see discussion in
\S 4.1).  While rotational velocities can be measured readily in these
gas-rich systems, obtaining accurate central stellar velocity dispersions for
them is extremely challenging because of the faintness or absence of a central
spheroidal component.  We can place a useful limit on the ``bulge'' velocity
dispersion of such systems by measuring the dispersion of its central nuclear
star cluster, which is commonly found (B\"oker et al.  2002).  These are
challenging observations, because the clusters are faint and the dispersions
are small, requiring echelle resolution on large telescopes.  We have been
able to locate suitable \sig\ measurements for a total of 10 late-type,
bulgeless spirals: NGC~598 (M33; Kormendy \& McClure 1993), NGC~4395
(Filippenko \& Ho 2003), and eight galaxies from the B\"oker et al. (2002)
survey (Walcher et al. 2005).
}

\item{{\it Sample 4 ---}\ \ \ Courteau et al. (2007) give the latest
compilation of galaxies having both \sig\ and \vm\ determined largely from
extended optical rotation curves [59 of their objects come from the study of
Prugniel et al. (2001), whose measurements of \vm\ were derived from \hi\
line widths].  Of the 164 galaxies in Courteau's compilation, 54
are not included in our samples 1--3; these form the last sample in our
study.  We distinguish three subsets: (a) 21 galaxies with \sig\ taken from
Courteau et al. and 

\vskip 0.3cm
\begin{figure*}[t]
\centerline{\psfig{file=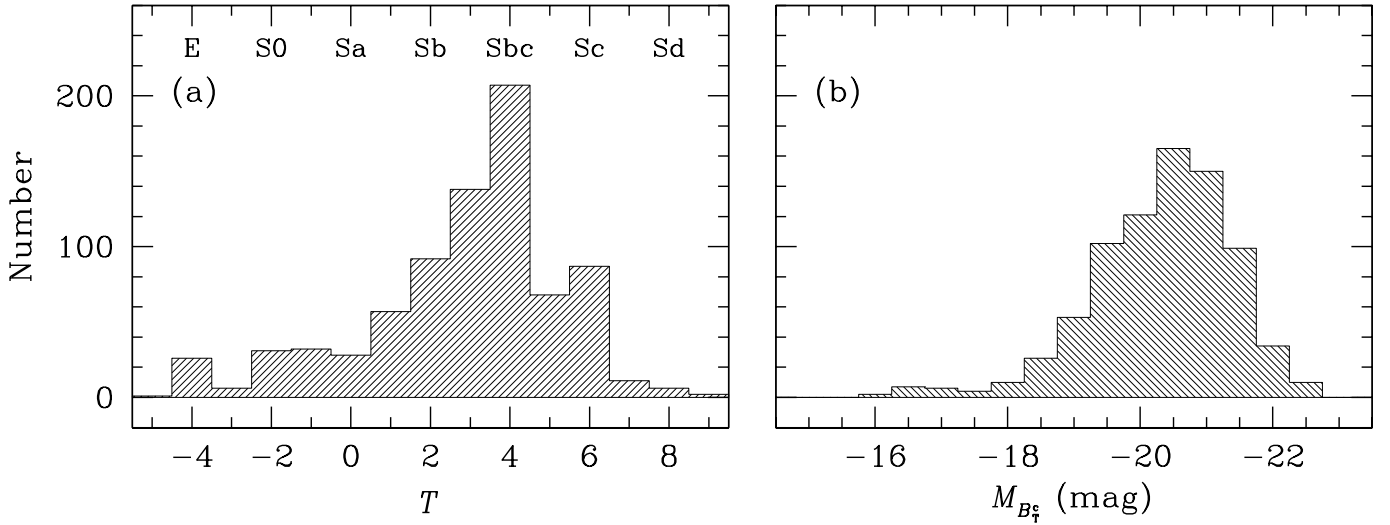,width=19.5cm,angle=-90}}
\figcaption[fig1.ps]{Distribution of
({\it a}) morphological type index $T$ and ({\it b}) total $B$-band absolute
magnitude, corrected for Galactic and internal extinction, for the 792
galaxies in our sample.  A rough mapping between $T$ and Hubble type is
given on the upper abscissa of panel ({\it a}).
\label{fig1}}
\end{figure*}
\vskip 0.3cm

\noindent
\vm\ derived from \hi\ line widths in Hyperleda; (b) 26
galaxies with \vm\ taken from Courteau et al. and \sig\ taken from
Hyperleda; and (c) 7 galaxies with \sig\ and \vm\ both taken from Courteau et
al.  In the interest of maximizing the homogeneity of our combined sample, we
have decided to use \hi\ line widths whenever available even if Courteau et al.
give \vm\ from resolved rotation curves.  Similarly, we give preference to
the \sig\ values in Hyperleda because these represent weighted averages of all
available literature measurements, and they have been scaled to a common
system adopted for the rest of our study.  The velocity dispersions that 
overlap between Courteau and Hyperleda show excellent agreement, with 
Courteau's values on average 3\% higher than Hyperleda's.
}
\end{itemize}

The combination of the above four samples produces a final compilation of 792 
galaxies, summarized in Figure~1 and Tables~1 and 2.  Most of the galaxies are 
nearby (median distance 23 Mpc) and luminous (median $M_{B_T^{\rm c}} = -20.3$ 
mag), spanning the entire range of Hubble types, from giant ellipticals to 
late-type spirals (morphological type index $T=-5$ to 8). 

The published \hi\ line widths have been corrected for instrumental resolution 
but not for ($1+z$) cosmological stretching or for broadening by interstellar 
turbulence.  A number of authors have discussed the impact of turbulence on
the line width measurements (e.g., Tully \& Fouqu\'e 1985; Fouqu\'e et al. 
1990).  Given the demographic make up of most of our sample (Fig.~1{\it b}), we
adopt the simple linear subtraction recommended by Bottinelli et al. (1983) 
for giant galaxies, 

\begin{equation}
W = W_{\rm obs} - W_{\rm turb}, 
\end{equation}

\noindent
where $W_{\rm obs}$ is the observed line width (at either 20\% or 50\% of the 
maximum), and the turbulent velocity is taken to be $W_{\rm turb} = 22$ \kms\ 
for $W_{20}$ and $W_{\rm turb} = 5$ \kms\ for $W_{50}$ (Verheijen \& Sancisi 
2001).

As the \hi\ line widths must be deprojected along the line-of-sight, we need to
pay careful attention to the adopted inclination angle.  Although Hyperleda 
conveniently lists inclination angles for our sample, they differ 
systematically and sometimes dramatically from the values given in other 
catalogs (see Appendix).  We therefore resorted to recompiling our own axial 
ratios and recalculating the inclination angles for the entire sample (Table~1).
We derive the inclination angle $i$ using Hubble's (1926) formula, 

\begin{equation}
{\rm cos}^2  i = {{q^2 - q_0^2}\over{1-q_0^2}},
\end{equation}

\noindent
which makes use of the apparent flattening of the galaxy ($q\equiv b/a$) as 
measured from the ratio of its semi-minor and semi-major isophotal diameters at 
a surface brightness of $\mu_B$ = 25 mag arcsec$^{-2}$, assuming that the 
intrinsic thickness of the disk ($q_0$) depends on morphological type as given 
in Paturel et al. (1997).  To mitigate against large uncertainties inherent in 
determining the inclination angle of nearly face-on systems, throughout this 
paper we remove galaxies that have inclination angles of $i < 30$\deg; these 
account for $\sim 10$\% of the original sample.  An important 
caveat, however, is that for most of the galaxies in our study we have 
absolutely {\it no}\ information on the spatial distribution of the neutral 
hydrogen.  Given that some \hi\ disks can be misaligned with respect to the 
stellar distribution (e.g., Lewis 1987), our optically derived inclination 
correction may not be valid in some instances.  Another source of uncertainty 
comes from strong warps and other types of nonaxisymmetric distortions in the 
\hi\ disk, which are known to be present in many disk galaxies (e.g., Baldwin 
et al. 1980; Bosma 1981; Lewis 1987; Richter \& Sancisi 1994; Haynes et al. 
1998). Indeed, in \S 4.3 we present evidence based on our sample that a 
sizable fraction of nearby galaxies may have strongly disturbed or dynamically 
unrelaxed \hi\ distributions.

Lastly, we need an empirical prescription to convert the corrected \hi\ line 
width into the maximum rotation velocity.  This issue has been investigated 
in a number of studies (e.g., Mathewson et al. 1992; Courteau 1997), the 
latest and most thorough being that of Paturel et al. (2003b), whose 
calibrations we adopt.  In detail, the calibrations depend on the resolution
of the 

\vskip 0.3cm
\begin{figure*}[t]
\centerline{\psfig{file=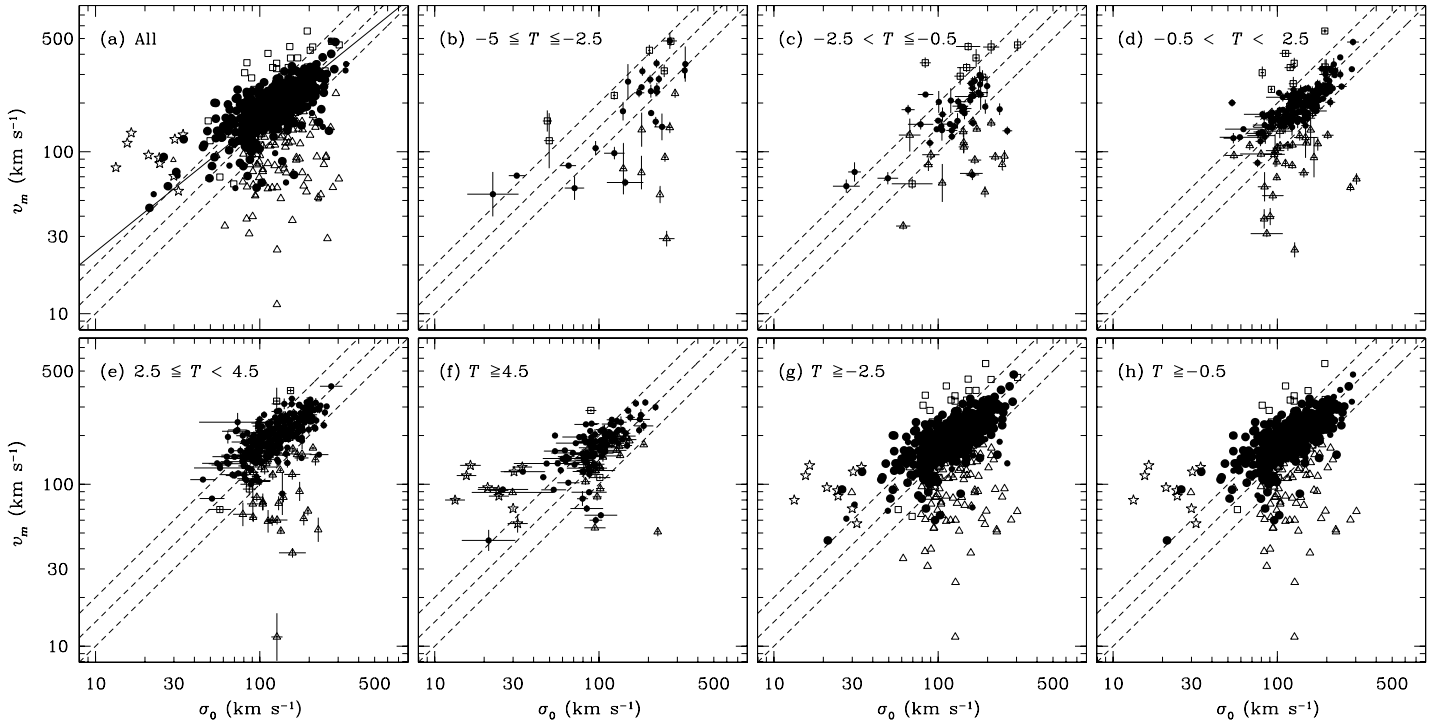,width=19.5cm,angle=270}}
\figcaption[fig2.ps]{
The relationship between central stellar velocity dispersion and the maximum
rotation velocity for galaxies of the following Hubble types:
({\it a}) all,
({\it b}) ellipticals ($-5 \leq T \leq -2.5$),
({\it c}) S0 ($-2.5 < T \leq -0.5$),
({\it d}) Sa--Sb ($-0.5 < T < 2.5$),
({\it e}) Sbc--Sc ($2.5 \leq T < 4.5$),
({\it f}) later than Sc ($T \geq 4.5$),
({\it g}) all disks, and
({\it h}) all spirals.
Sources that are low-velocity and high-velocity outliers in the Tully-Fisher
relation (Fig. 3) are plotted as open triangles and squares,
respectively.  The solid points belong to the ``kinematically normal''
sample.  In panels ({\it a}) and ({\it f})--({\it h}), the extreme late-type,
bulgeless spirals are plotted as stars.  For the sake of clarity,
error bars are not shown in panels ({\it a}), ({\it g}), and ({\it h}).  The
{\it dashed}\ lines denote, from bottom to top, \vm\ = \sig,
$\sqrt{2}\ \sigma_0$, and $\sqrt{4}\ \sigma_0$.  The best fit to the entire
sample of kinematically normal galaxies is plotted as a solid line in panel
({\it a}).
\label{fig2}}
\end{figure*}
\vskip 0.0cm

\noindent
observations, but for simplicity we just use those obtained for the 
highest resolution (8 \kms):

\begin{eqnarray}
\log 2\upsilon_m {\rm sin}i=(1.071\pm0.009)\log W_{50}-(0.210\pm0.023)
\nonumber
\end{eqnarray}

\begin{eqnarray}
\log 2\upsilon_m {\rm sin}i=(1.187\pm0.002)\log W_{20}-(0.543\pm0.005).
\end{eqnarray}

\vskip 0.3cm

\noindent
The uncertainty on \vm\ is taken to be the value formally given in 
Hyperleda.  This is a conservative estimate because in many cases the 
Hyperleda value incorporates a larger spread of individual measurements than 
we actually used.  For our sample, the values of \vm\ have an average 
uncertainty of 6.3\%, with a standard deviation of 5.4\%.

\vskip 0.3cm
\section{The \vm--\sig\ Relation}

Figure~2 summarizes our principal findings. Taken collectively (Fig.~2{\it a}),
the distribution of \vm\ versus \sig\ shows, at best, a loose correlation. 
Although the best-fit relations of Baes et al. (2003) or Pizzella et al. 
(2005) roughly bisect the cloud of points, the scatter is enormous, far 
greater than can be attributed to observational errors or potential 
sources of systematic uncertainty (e.g., inclination corrections for \vm\
or aperture corrections for \sig).  Separating the sample by Hubble type 
(Fig.~2{\it b}--2{\it f}) reveals three main culprits for the large scatter: 
(1) a systematic shift of zeropoint as a function of Hubble type, most clearly 
seen in the locus of the ridgeline defining the upper envelope of the 
distribution of points; (2) the existence of subset of low-\sig\ (\sig\ \lax\ 
50 \kms), extreme late-type galaxies ($T \approx 6-9$, Hubble types Scd--Sdm) 
that have almost constant \vm\ ($\sim 100$ \kms); and (3) a cloud of points, 
present in all Hubble type bins, but especially prominent for earlier-type 
systems, with very low values of \vm/\sig\ (plotted as {\it open triangles}).  

To assess how much of the scatter in the \vm--\sig\ diagram is intrinsic, we 
make use of the Tully-Fisher relation to constrain what value of \vm\ any 
particular galaxy {\it ought}\ to have given its luminosity.  Since the Hubble 
type mix of a sample affects the slope, normalization, and scatter of the 
Tully-Fisher relation (e.g., Roberts 1978; Rubin et al. 1985; Bell \& de~Jong 
2001; De~Rijcke et al. 2007; Pizagno et al. 2007), and the variations are 
minimized in the near-infrared (e.g., Verheijen 2001), we assembled $K_s$-band 
(2.16 \micron) magnitudes for nearly the entire sample using the Extended 
Source Catalog of the Two-Micron All-Sky Survey (2MASS; Skrutskie et al. 
2006).  The photometry pertains to the ``total'' magnitudes (Table~1), and for
simplicity we do not correct for Galactic or internal extinction, which should
be quite small in the $K_s$ band.  Figure~3 shows that a $K_s$-band 
Tully-Fisher relation exists for all Hubble types, including E and S0, and, 
importantly in the present context, that there is little obvious variation in 
the slope, normalization, or scatter of the relation across the wide range of 
Hubble types included in our sample.  Overplotted on the figure is the fit 
derived by Verheijen (2001) for the $K^\prime$ band, which is quite similar to 
the 2MASS $K_s$ band (Bessel 2005).  While the observed scatter of our 
Tully-Fisher relation is larger than that found by Verheijen (2001), a result 
that can be anticipated considering the larger distance errors and greater 
heterogeneity of our sample, most of our objects (625/792 or 70\%; {\it solid 
points}) fall comfortably within the boundaries that enclose twice the rms 
scatter

\vskip 0.3cm
\begin{figure*}[t]
\centerline{\psfig{file=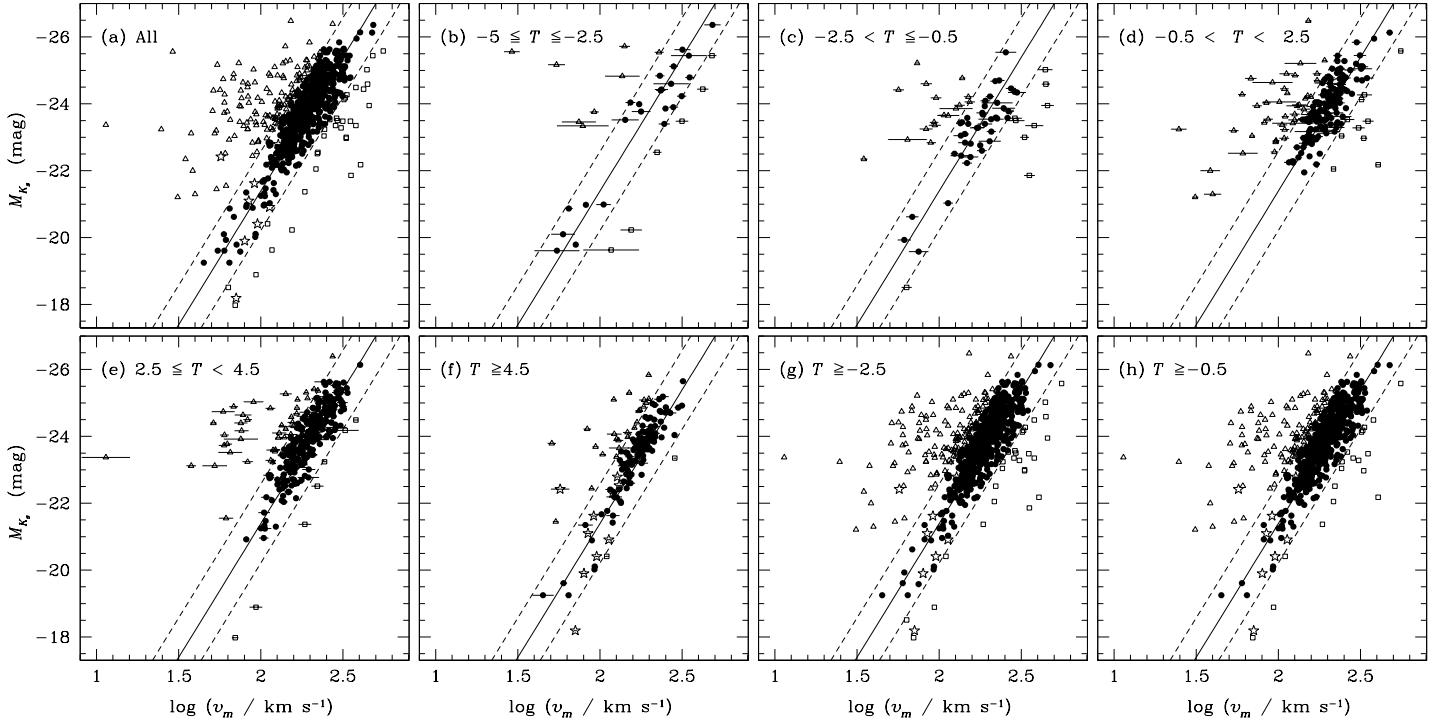,width=19.5cm,angle=270}}
\figcaption[fig3.ps]{
Tully-Fisher relation in the $K_s$ band for galaxies of the following Hubble
types:
({\it a}) all,
({\it b}) ellipticals ($-5 \leq T \leq -2.5$),
({\it c}) S0 ($-2.5 < T \leq -0.5$),
({\it d}) Sa--Sb ($-0.5 < T < 2.5$),
({\it e}) Sbc--Sc ($2.5 \leq T < 4.5$),
({\it f}) later than Sc ($T \geq 4.5$),
({\it g}) all disks, and
({\it h}) all spirals.
The {\it solid}\ line represents the $K^\prime$-band Tully-Fisher relation
from Verheijen (2001); the {\it dashed}\ lines mark the region that has twice
the rms scatter.  Sources that are low-velocity and high-velocity outliers
are plotted as open triangles and squares, respectively; the solid points
are considered to be the ``kinematically normal'' members of the sample.
In panels ({\it a}) and ({\it f})--({\it h}), the extreme late-type,
bulgeless spirals are plotted as stars.  For the sake of clarity,
error bars are not shown in panels ({\it a}), ({\it g}), and ({\it h}).
\label{fig3}}
\end{figure*}
\vskip 0.3cm

\noindent
of Verheijen's ``\hi'' sample (0.59 mag); hereafter we will refer to 
these as the ``kinematically normal'' objects.  Nevertheless, Figure~3 shows 
that there are a significant number of outliers, which we loosely and somewhat 
arbitrarily define to be those that lie outside of the 2-rms band.  Closer 
inspection reveals that there is an {\it excess}\ of low-velocity objects 
(132/792 or 17\%; {\it open triangles}) compared to high-velocity ones (35/792 
or 4\%; {\it open squares}).  This is most obvious among the spirals ($T \geq 
-0.5$) as a low-\vm\ ``plume,'' but to a lesser extent it can be 
seen also among the E and S0 systems.

Within this backdrop, the distribution of points in the \vm--\sig\ diagram can 
now be more easily interpreted.  The ``kinematically normal'' objects on the 
Tully-Fisher relation obey a loose correlation between \vm\ and \sig, roughly 
occupying the region from \vm\ = \sig\ to \vm\ = $\sqrt{4}$\sig, depending on 
Hubble type.  The correlation has significant scatter, and it is nonlinear.  
An ordinary least-squares bisector fit for the entire sample (Fig.~2{\it a}) 
yields 

\begin{equation}
\log \upsilon_m = (0.82\pm0.027) \log \sigma_0 + (0.57\pm0.058),
\end{equation}

\vskip 0.2cm
\noindent
very similar to the fit reported by Ferrarese (2002).  There is no significant 
variation of the slope with Hubble type.  Limiting the fit to the 550 
kinematically normal spiral galaxies, the fit is nearly identical:

\begin{equation}
\log \upsilon_m = (0.80\pm0.029) \log \sigma_0 + (0.62\pm0.062).
\end{equation}

\noindent
The bulgeless late-type galaxies, 
which are largely kinematically normal in the Tully-Fisher relation, continue 
to depart notably from the rest of the sample in the \vm--\sig\ diagram.  It 
is of interest to note that the few late-type spirals (NGC~6140, NGC~6689, 
PGC~28990) that do host small bulges (as opposed to nuclear star clusters), 
along with some dwarf ellipticals and S0s (NGC~3870, NGC~7077, PGC~5441, 
PGC~71938), still follow the low-velocity extrapolation of the \vm--\sig\ 
relation.  Within the large scatter, we find no compelling evidence for a 
break in the \vm--\sig\ relation at low velocities.  Objects with low 
\vm/\sig\ (\lax 1) comprise almost exclusively the low-velocity outliers in 
the Tully-Fisher relation.  The nature of the low-\vm/\sig\ objects is 
discussed further in \S 4.3.  The high-velocity outliers, on the other hand, 
occupy the upper envelope of the \vm--\sig\ relation, especially for 
\sig\ \gax\ 80 \kms.  Had these objects not been excluded, they would bias the 
\vm--\sig\ relation to a steeper slope.

Figure~4 highlights the trends with Hubble type more explicitly, showing the 
distribution of \vm/\sig\ for six bins of morphological types.  Although the 
binning of the morphological types is somewhat arbitrary, nevertheless the 
pattern is clear: as the Hubble type becomes later, the median of the 
distribution of \vm/\sig\ systematically shifts to larger values, from 
$\sim$1.2 for ellipticals, to $\sim$1.4 for early-type spirals, to $\sim$1.8 
for late-type spirals, and, very dramatically, to $\sim$4.6 for extreme 
late-type, bulgeless spirals.  This trend was already largely noticed by 
earlier studies (Whitmore et al. 1979; Whitmore \& Kirshner 1981; Franx 1993;
Whittle 1992a; Zasov et al. 2005).  Since we do not have reliable 
bulge-to-disk photometric decompositions for the majority of our sample, we 
constructed a surrogate measure of bulge luminosity for the disk galaxies by 
using the observed values of \sig\ in combination with the Faber-Jackson 
(1976) relation.  Using the 1072 elliptical ($T \leq -2.5$) galaxies with 
measurements of \sig\ listed in Hyperleda, we find the following 

\vskip 1.3cm
\psfig{file=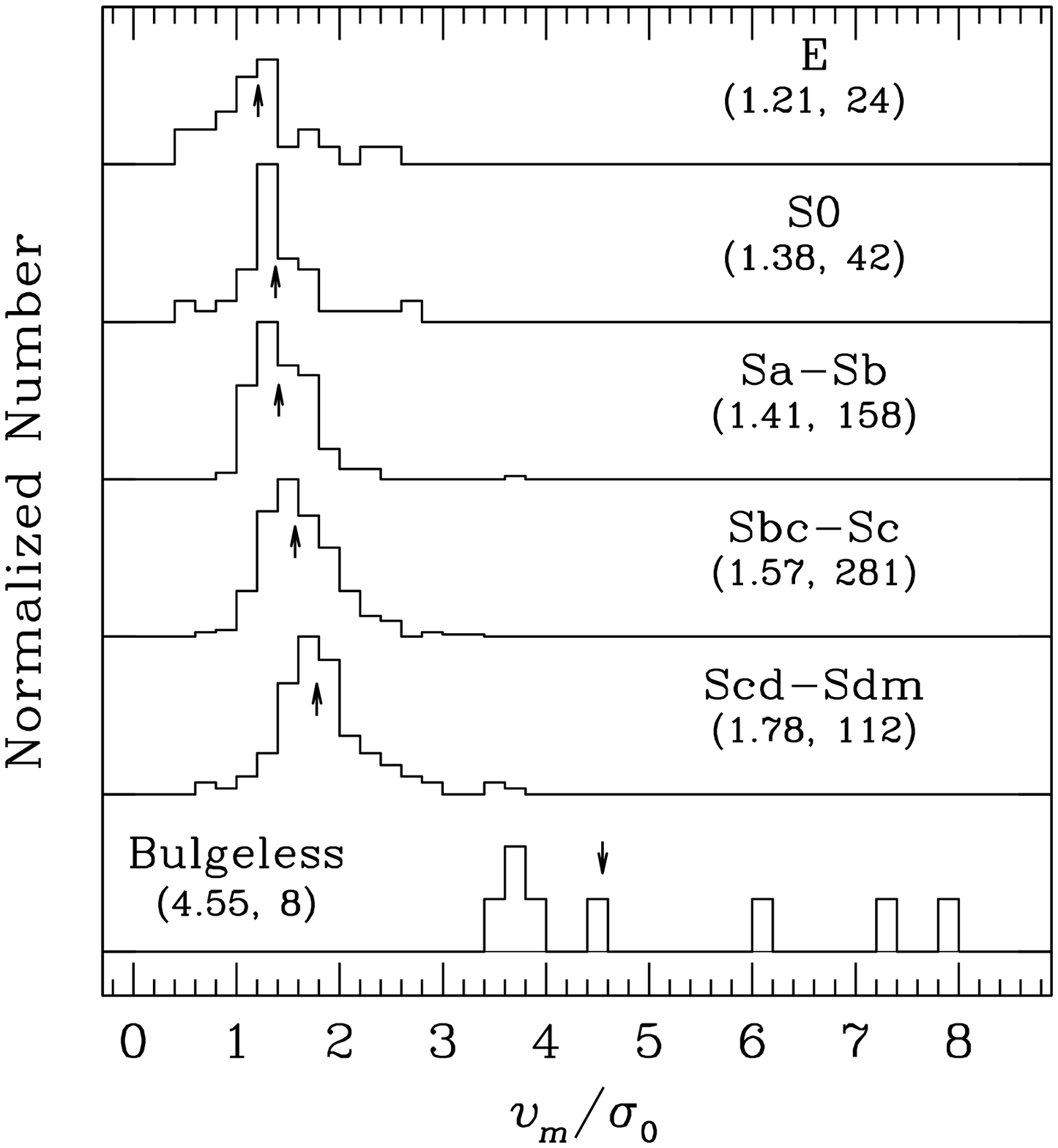,width=8.5cm,angle=0}
\figcaption[fig4.ps]{The distribution of the ratio of disk rotation velocity
to bulge velocity dispersion for the ``kinematically normal'' galaxies
(see \S 3) of Hubble type E, S0, Sa--Sb, Sbc--Sc, Scd--Sdm, and extreme
late-type, bulgeless spirals.  The median of each distribution is given,
followed by the number of galaxies in the group; the median is also marked by
an arrow.
\label{fig4}}
\vskip 0.3cm

\noindent
ordinary
least-squares bisector fit\footnote{This fit is roughly consistent with the
SDSS results of Bernardi et al.  (2003).  For example, for a velocity
dispersion of \sig\ = 200 \kms, their equation (11) predicts $M_g = -19.33$
mag, which, for $g-B = -0.53$ mag expected for elliptical galaxies (Fukugita
et al. 1995), translates to $M_B = -18.80$ mag.  Our fit gives $M_B = -18.49$
mag.}: $M_{B_{\rm T}}=-6.80\log \sigma_0-4.89$.
This method of estimating the ``bulge'' luminosity certainly has limitations.
The Faber-Jackson relation for bulges is, as yet, not well-determined: it
appears to differ systematically with Hubble type (Whitmore \& Kirshner 1981;
Kormendy \& 

\vskip 0.3cm
\psfig{file=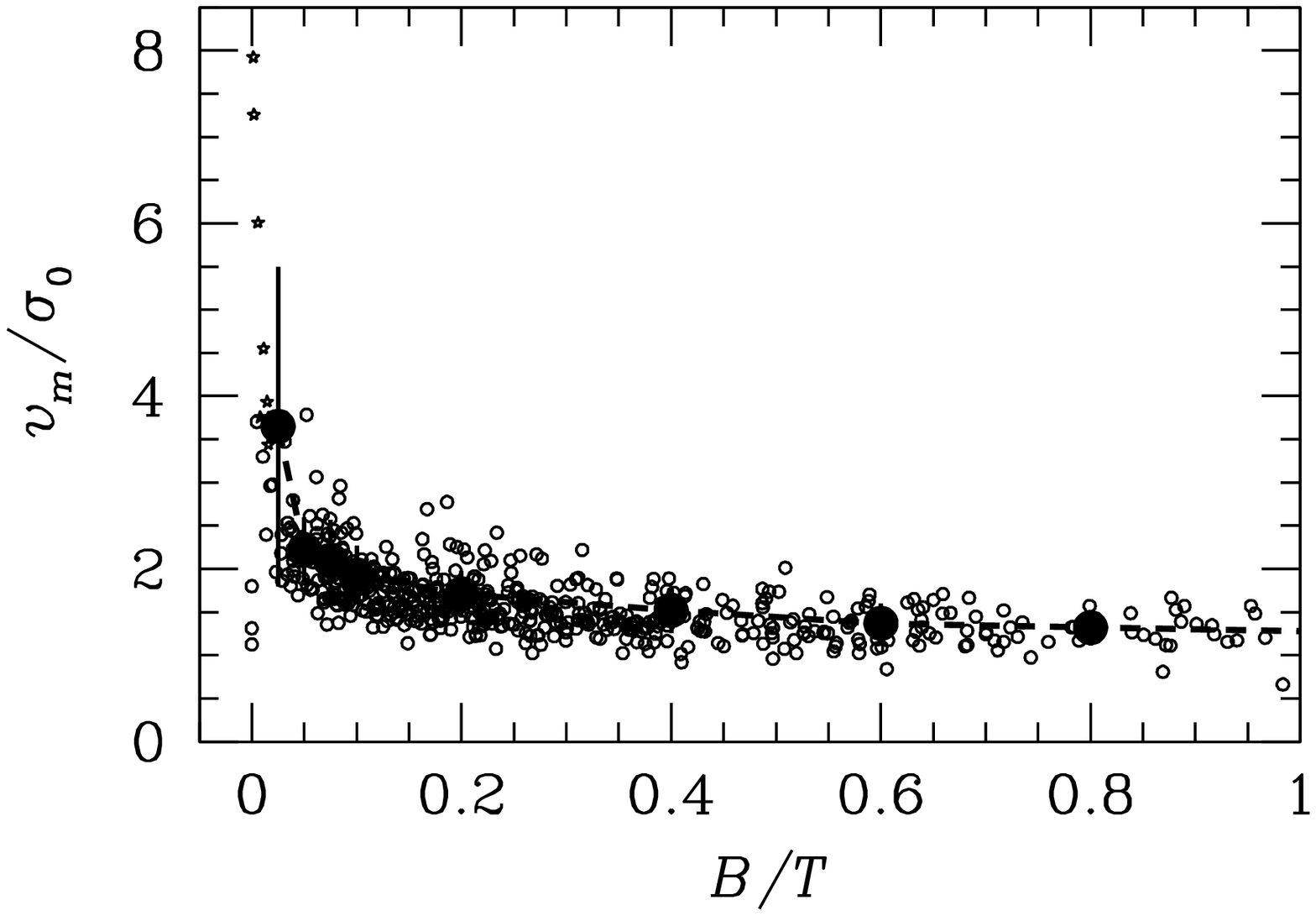,width=8.5cm,angle=0}
\figcaption[fig5.ps]{The variation of the ratio of disk rotation
velocity to bulge velocity dispersion as a function of the bulge-to-total
luminosity ratio.  Here the ``bulge'' luminosity is calculated from the
Faber-Jackson relation, as determined from a sample of 1072 elliptical
galaxies in Hyperleda.  The extreme late-type, bulgeless spirals are plotted
as stars. The solid points connected by the dashed line show the mean and
standard deviation of \vm/\sig\ binned by $B/T$ for the ``kinematically
normal'' sources (see \S 3).
\label{fig5}}
\vskip 0.3cm

\noindent
Illingworth 1983; but see Whittle 1992b) as well as between 
classical bulges and ``pseudo-bulges'' (Kormendy \& Kennicutt 2004), and it 
shows significant intrinsic scatter within each Hubble type.  Nevertheless, it 
is debatable whether these sources of systematic uncertainty are any more 
serious than the vagaries of photometric bulge-to-disk decomposition, 
especially for such a large and diverse sample of galaxies.  With these 
caveats in mind, Figure~5 shows that \vm/\sig\ clearly varies systematically 
with bulge-to-total ($B/T$) luminosity ratio\footnote{A small fraction of the 
sample, not shown in Figure 5, has $B/T$ values that formally exceed unity. 
This should not be viewed as too alarming, in view of the indirect method by 
which we have estimated the ``bulge'' luminosity.}.  A given value of $B/T$ 
can host a large range of \vm/\sig, but on average \vm/\sig\ increases as the 
prominence of the bulge component decreases, qualitatively consistent with the 
trends noted in Figures~2 and 4. 

Another crude, but more direct, method to estimate the degree of bulge 
dominance is to calculate the ``concentration index'' of the light profile, 
which is roughly related to the bulge-to-disk ratio or Hubble type  (e.g., 
Doi et al. 1993).  The correlation between concentration index and Hubble type 
in the SDSS database has been investigated by Shimasaku et al. (2001) and 
Strateva et al. (2001).  This is the approach taken by Courteau et al. (2007), 
who defined a concentration index $C_{28}\equiv 5 \log (r_{80}/r_{20})$, where 
$r_{20}$ and $r_{80}$ are the radii that enclose 20\% and 80\% of the total 
light, respectively.  They computed $C_{28}$ for 81 out of the 164 galaxies 
in their sample with available $i$-band images in the SDSS.  We follow 
Courteau et al. in using the concentration index as a surrogate indicator for 
the density distribution of the galaxy light profile, but instead of $C_{28}$, 
we simply adopt the cataloged $i$-band Petrosian radii enclosing 50\% 
($r_{\rm P50}$) and 90\% ($r_{\rm P90}$) of the light to form an equivalent 
concentration index $C_{59} \equiv 5\log(r_{\rm P90}/r_{\rm P50})$.  Not 
surprisingly, $C_{28}$ and $C_{59}$ are strongly correlated (Fig.~6{\it a}), 
albeit with significant scatter, which, too, is not unexpected.  
Figure~6{\it b}\ illustrates that substituting $C_{59}$ for $C_{28}$ 
faithfully recovers the correlation between \vm/\sig\ and concentration index 
presented in Courteau et al. (2007; see their Fig.~2).  If anything, using 
$C_{59}$ instead of $C_{28}$ seems to produce a diagram with even somewhat less 
scatter\footnote{Figure~6{\it b}\ contains fewer data points than Figure~2 of 
Courteau et al. (2007) because the latter treated repeat observations of the 
same objects as independent galaxies (S. Courteau 2006, private 
communications).}.

The majority of our sample (76\%) have photometric data available in 
the Fifth Data Release of SDSS (Adelman-McCarthy et al. 2007), and so we can 
calculate the $C_{59}$ parameter for these objects (Table~1).  As anticipated 
from the Courteau et al. study, Figure~7 illustrates that our sample indeed 
also behaves very similarly: a systematic trend exists between \vm/\sig\
and $C_{59}$, although the scatter is substantial.

The density distribution, as traced by the morphological type, bulge-to-disk 
ratio, or concentration index, all of which are loosely mutually correlated, 
appears to be the dominant factor that determines a galaxy's \vm/\sig\ ratio.  
We have examined the total galaxy luminosity as a possible additional 
parameter, but it has little or no effect.  This is not surprising, because 
with the exception of very late-type systems, a galaxy's total luminosity 
shows only subtle variation with Hubble type (e.g., Roberts \& Haynes 1994).  
A galaxy's integrated broad-band colors ($U-B$, $B-V$, $B-I$, $B-K_s$), to the 
extent that they are available from Hyperleda, do show a moderate correlation 
with \vm/\sig, but this is likely just a reflection of the dependence of galaxy 
color on 

\vskip 0.3cm
\psfig{file=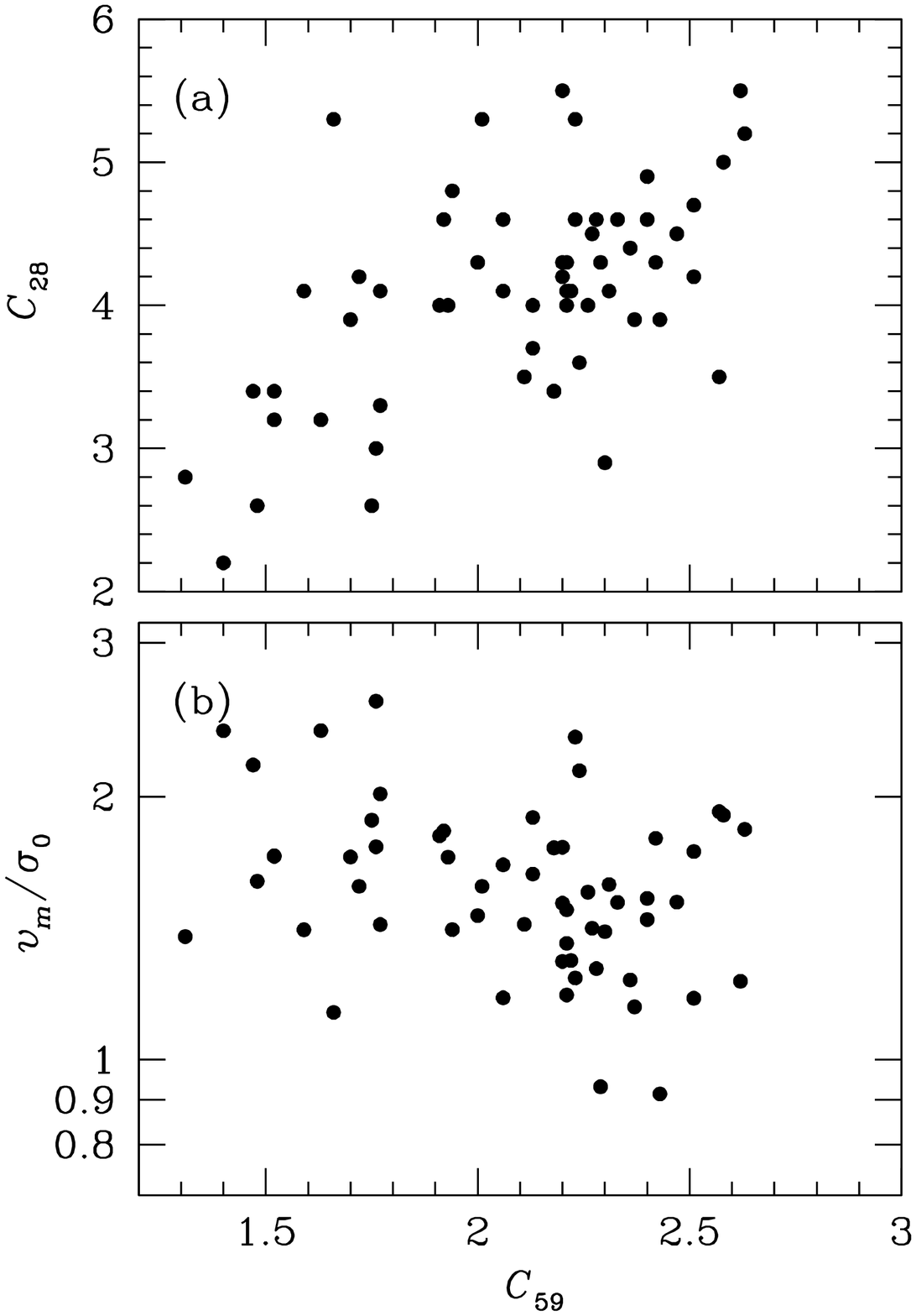,width=9.0cm,angle=0}
\figcaption[fig6.ps]{
({\it a}) The relation between $i$-band concentration indices as derived in
the Petrosian system of SDSS, $C_{59}$, and as given by Courteau et al. (2007),
$C_{28}$. The two concentration indices are strongly correlated.  ({\it b}) The
variation of \vm/\sig\ with $C_{59}$ for the sample of Courteau et al.,
plotted in the same fashion as Fig.~2 of Courteau et al., demonstrating that
$C_{59}$ can be substituted for $C_{28}$.
\label{fig6}}
\vskip 0.3cm

\noindent
morphological type or concentration.  

One caveat should be kept in mind.  It is possible that the central velocity 
dispersions of the later-type galaxies are underestimated due to contamination
by disk light.  If so, this could mimic the variation of \vm/\sig\ with 
galaxy type or bulge-to-disk ratio.  This possibility needs to be checked 
with detailed spatially resolved stellar kinematical data.

\section{Discussion}

\subsection{Implications for Black Hole Demographics}

This study was initiated with the intention of better quantifying the relation 
between disk rotation velocity and bulge stellar velocity dispersion, with the 
hope that such a correlation might be used as a new empirical tool to study the 
demographics of central black holes in galaxies.  Earlier studies (Ferrarese 
2002; Baes et al. 2003), based on relatively small samples with extended 
optical rotation curves, found that galaxies over a wide range of Hubble 
types, from mid-type spirals to giant ellipticals, obey a strong correlation 
between \vm\ and \sig\ with almost no intrinsic scatter, at least for \vm\ 
\gax\ 80 \kms.  Subsequent work has emphasized that the \vm--\sig\ relation 
may depend on surface brightness, with low-surface brightness galaxies lying 
systematically offset (larger \vm\ for a given \sig) from a still tight 
sequence defined by high-surface brightness galaxies (Pizzella et al. 2005; 
Buyle et al. 2006). If high-surface brightness galaxies truly do obey a tight 
\vm--\sig\ relation, one would have to 

\vskip 0.3cm
\psfig{file=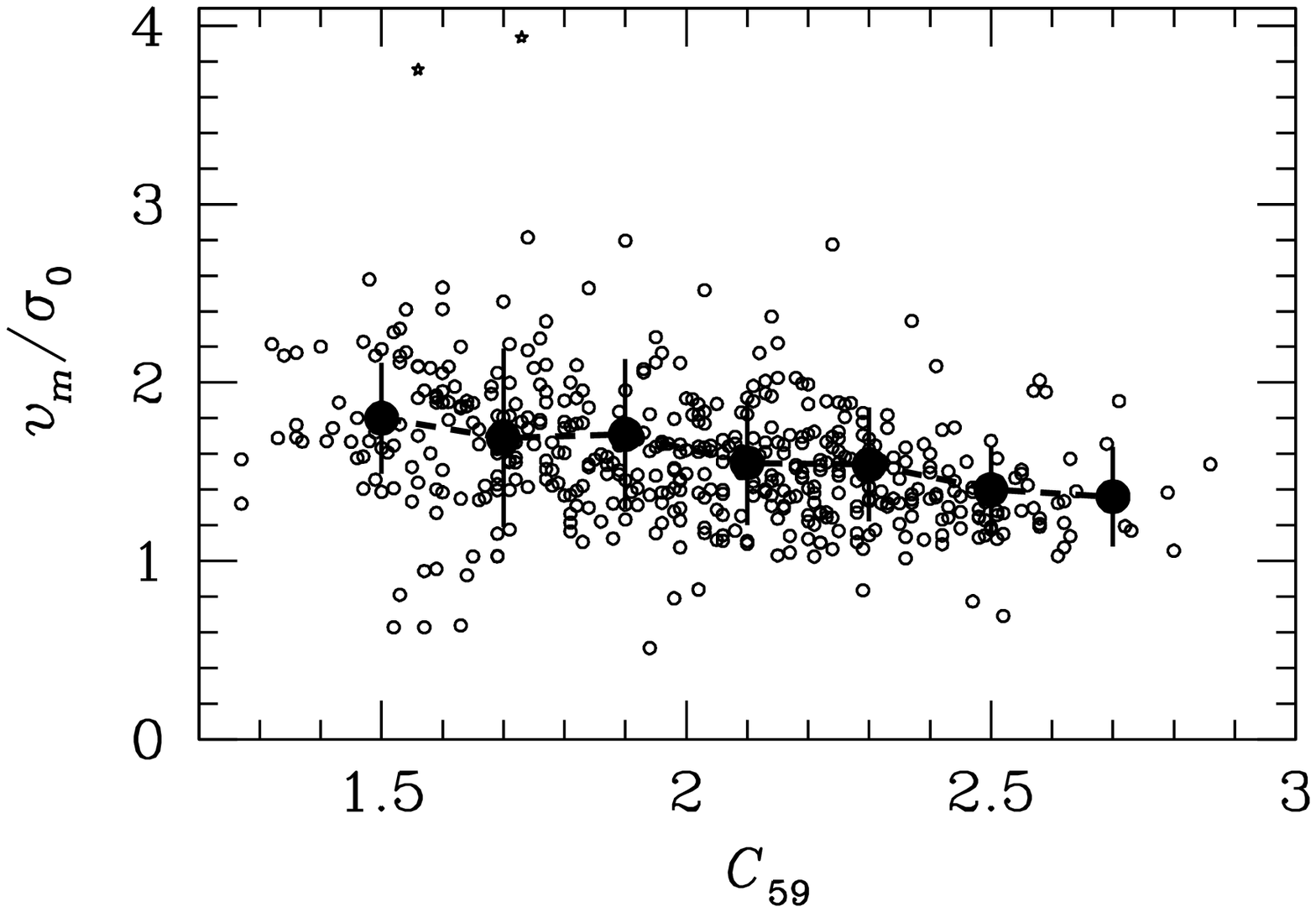,width=8.5cm,angle=0}
\figcaption[fig7.ps]{
The variation of \vm/\sig\ with the concentration index $C_{59}$.  The extreme
late-type, bulgeless spirals are plotted as stars.  The solid points connected
by the dashed line show the mean and standard deviation of \vm/\sig\ binned
by $C_{59}$, for the ``kinematically normal'' sources (see \S 3).
\label{fig7}}
\vskip 0.3cm

\noindent
reevaluate which dynamical variable 
(\sig\ or \vm) or, equivalently, which structural component of the galaxy 
(bulge or halo) is more fundamentally related to black hole mass.

Nevertheless, there is reason to suspect that the \vm--\sig\ relation may not 
be as simple as had been claimed.   Indeed, from the initial work of Whitmore 
and collaborators (Whitmore et al. 1979; Whitmore \& Kirshner 1981), as well 
as from subsequent updates thereof (Fall 1987; Whittle 1992a; Franx 1993; 
Nelson \& Whittle 1996), one could already have concluded that the ratio 
\vm/\sig\ is {\it not}\ a constant, but rather that it varies systematically 
as a function of bulge-to-disk ratio, in the sense that early-type galaxies 
have characteristically smaller \vm/\sig\ ratios than late-type galaxies.  It 
was also apparent that there is significant scatter in \vm/\sig\ for any given 
bulge-to-disk ratio.  Baes et al. (2004) speculated that the culprit for the 
increased scatter could lie in Whitmore's use of integrated \hi\ line widths, 
which may not accurately trace the flat part of the rotation curve.  This 
explanation, however, seems implausible in light of the long history of 
Tully-Fisher studies that have made use of integrated \hi\ line widths, not to 
mention of the many comparisons made with optically derived rotation curves 
(e.g., Rubin et al. 1978; Thonnard 1983; Mathewson et al. 1992; Courteau 1997; 
Paturel et al. 2003b).  Our own comparison shown in Figure~9, though limited, 
reinforces previous conclusions that integrated \hi\ line widths serve as an 
effective proxy for rotation velocities measured from extended rotation 
curves, at least for spiral galaxies.  Moreover, even if one were to admit 
that \hi-based rotation velocities were less accurate, one would still have to 
explain why \vm/\sig\ changes systematically with bulge-to-disk ratio.  Zasov 
et al. (2005) provided further evidence that the \vm--\sig\ relation contains 
significant intrinsic scatter.  Using a collection of 41 galaxies with black 
hole masses, central stellar velocity dispersions, and rotational velocities 
from {\it optical}\ rotation curves, these authors concluded that although 
\mbh\ does correlate with \vm, it is better correlated with \sig.  They 
also find that, at a given \vm, S0--Sab galaxies tend to have larger \mbh\ 
than later-type galaxies.  From this, one can infer that earlier-type galaxies 
possess a smaller \vm/\sig\ ratio than later-type galaxies.  The latest 
demonstration that \vm/\sig\ varies considerably and systematically with 
galaxy type comes from Courteau et al. (2007), whose \vm\ compilation---now 
increased to 164 galaxies---are also based largely on spatially resolved 
optical rotation curves.  Their study shows that the \vm--\sig\ relation 
exhibits marked scatter and that \vm/\sig\ decreases steadily, albeit with 
large fluctuations, with increasing galaxy light concentration.  

The present study evaluates the \vm--\sig\ relation with the largest sample to 
date, totaling 792 galaxies.   This is made possible not only by access to a 
large number of new \sig\ measurements from SDSS, but also by taking advantage 
of integrated \hi\ line widths to estimate rotation velocities, a shortcut 
whose efficacy has been amply proved.  As anticipated, we find that the 
\vm--\sig\ relation is far from tight. The zeropoint shifts systematically as 
a function of Hubble type, and within each Hubble type bin the scatter is at 
least a factor of 2--3.  From the point of view of using \vm\ to predict \mbh, 
this is not promising, not if other properties of the host galaxy such as 
bulge luminosity (Kormendy \& Gebhardt 2001; Marconi \& Hunt 2003), and 
especially \sig\ (Tremaine et al. 2002), are available.  On the other hand, 
there are instances when the stellar component of the host galaxy can be 
extraordinarily challenging, if not impossible, to detect, either 
kinematically or photometrically.  This is often the case in luminous active 
galactic nuclei or quasars.  The host galaxy's neutral hydrogen gas, by 
contrast, being immune to the bright glare of the active nucleus, may be more 
readily detectable with deep radio observations. Under these circumstances, an 
integrated line width, which can be extracted from an \hi\ spectrum of even 
moderate signal-to-noise ratio, may be the {\it only}\ constraint available on 
the host galaxy.   The recent analysis by Ho et al. (2007b) of a new \hi\ 
survey of active galaxies (Ho et al. 2007a) demonstrates the rich body of 
information on the host galaxy that can be ascertained in this unique fashion.
The existence of low-\vm/\sig\ outliers (\S 4.3) certainly complicates 
matters, but this difficulty is not insurmountable because these objects 
can be recognized as outliers in the Tully-Fisher relation, if an independent 
estimate of the total host galaxy luminosity can be estimated.  Alternatively, 
Ho et al. (2007b) suggest that this kinematically anomalous population may 
have a greater tendency to exhibit single-peaked and/or highly asymmetric 
line profiles.  While this trend still needs to be verified with better data, 
it may offer a very effective, practical strategy to weed out unwanted 
contaminants.

It is important to stress that the large scatter of the \vm--\sig\ relation is 
not a consequence of mixing low- and high-surface brightness galaxies.  
Although we cannot rigorously identify low-surface brightness galaxies in our 
sample with the material at hand, given the very bright luminosities of 
most of our sample it is highly improbable that a large fraction of them 
are low-surface brightness galaxies (see, e.g., Sprayberry et al. 1997).

The absence of a single, universal \vm--\sig\ correlation, in 
concert with the general consensus that the \mbh--\sig\ relation {\it is}\ 
tight, casts doubt on Ferrarese's (2002) hypothesis that the halo mass is more 
fundamentally connected to the black hole mass than is the bulge mass. Judging 
from both direct dynamical searches of black holes (e.g., Magorrian et al. 
1998; Kormendy 2004, and references therein) and the high detection rate of 
active nuclei in early-type galaxies (Ho et al. 1997; Ho 2004), the occupation
fraction of black holes in bulges is very high, perhaps approaching unity.  By 
contrast, the inverse may be true in late-type and dwarf galaxies.  None 
has so far yielded a direct dynamical detection of a central black hole 
(Gebhardt et al. 2001; Merritt et al. 2001; Valluri et al. 2005), and nuclear 
activity is exceedingly rare (Ho et al. 1997; Ulvestad \& Ho 2002), though not 
unknown\footnote{Previous authors (Ferrarese 2002; Baes et al. 2003; Buyle et 
al. 2006) have used the apparent break down of the \vm--\sig\ relation at low 
velocities to infer that low-mass galaxies do not host central black holes.
This is incorrect. Bona fide examples of active galactic nuclei with low-mass 
black holes in low-mass (even dwarf) galaxies {\it are}\ known.  Although 
such objects are apparently rare, current search techniques introduce 
strong selection biases and preclude any definitive conclusions regarding
their true space densities (Greene \& Ho 2007a).} (Filippenko \& Ho 2003; 
Barth et al. 2004, 2005; Greene \& Ho 2004, 2007a, 2007b).  
Thus, with a few exceptions, central black holes, be they active or inactive,  
are associated almost exclusively with bulges, {\it not}\ halos.  The 
very late-type spirals included in this study (sample 3) possess substantial 
rotation velocities ($65 < \upsilon_m < 140$ \kms) and correspondingly 
non-negligible halo masses but effectively no bulges (B\"oker et al. 2003): 
their central stellar velocity dispersion, $13 < \sigma_0 < 34$ \kms, arises 
not from a bulge but a compact, nuclear star cluster (Kormendy \& McClure 
1993; Filippenko \& Ho 2003; Walcher et al. 2005).  If \mbh\ were more 
fundamentally linked with \vm\ than \sig, we would expect these galaxies to 
possess black holes with \mbh\ $\approx 2\times 10^{5}$ to $4\times 10^{6}$ 
\solmass\ [using the \vm--\sig\ relation of Baes et al. (2003) and the 
\mbh--\sig\ relation of Tremaine et al. (2002)], and yet none but NGC~4395 is 
known for sure to contain a central black hole (Filippenko \& Ho 2003).  M33 
(NGC~598) has the most stringent limit: \mbh\ $< 1500$ \solmass\ (Gebhardt et 
al.  2001).
 
\vskip 0.8cm
\subsection{Implications for Galaxy Formation}

Discouraging as it may be as a black hole mass predictor, the \vm--\sig\ 
relation can be regarded as a useful, additional constraint on galaxy 
formation models.  What factors determine the \vm/\sig\ ratio in any given 
galaxy?  Our study, in agreement with that of Courteau et al.  (2007), shows 
that \vm/\sig\ depends, at least in part, on the mass distribution within a 
galaxy.  This is evident in the correlation between \vm/\sig\ and three 
mutually related quantities that reflect the density profile of a galaxy, 
namely its Hubble type, bulge-to-disk ratio, and concentration index.  The 
trends are systematic and statistically significant, but none of the 
correlations can be regarded as particularly tight, suggesting that 
\vm/\sig\ probably depends on more than just the density profile alone.  This
can be anticipated from consideration of the collisionless Boltzmann equation 
for a stellar system in gravitational equilibrium, in which the rotation 
velocity and velocity dispersion at radius $r$ are related by

\begin{equation}
\upsilon^2(r) = \sigma_r^2(r) 
\left[-{{d{\rm ln}\rho(r)}\over{d{\rm ln}r}} -
{{d{\rm ln}\sigma_r^2(r)}\over{d{\rm ln}r}} -
2\left(1 - {{\sigma_\theta^2}\over{\sigma_r^2}}\right) \right].
\end{equation}

\vskip 0.3cm
\noindent
Here $\sigma_r$ and $\sigma_\theta$ are the velocity dispersions along the 
radial and tangential directions, and $\rho$ is the stellar density profile.
Following Courteau et al. (2007), we can recast this equation as 

\begin{equation}
{{\upsilon_m}\over{\sigma_0}} = {{\upsilon_m}\over{\upsilon_0}} 
\left[-{{d{\rm ln}\rho(r)}\over{d{\rm ln}r}} -
{{d{\rm ln}\sigma_r^2(r)}\over{d{\rm ln}r}} -
2\left(1 - {{\sigma_\theta^2}\over{\sigma_r^2}}\right) \right]^{1/2},
\end{equation}

\vskip 0.3cm
\noindent
where $\upsilon_0$ denotes the rotation velocity in the inner part of the 
galaxy.  The dependence of \vm/\sig\ on the density profile enters directly 
through the first term in the brackets as well as through 
$\upsilon_m/\upsilon_0$, which depends on the shape of the rotation curve and 
hence on the detailed mass distribution of the galaxy (Rubin et al. 1978, 
1985).  

The challenge for galaxy formation models is not to explain why \vm/\sig\ 
varies (as it is obvious from equation 7 that it should), but rather why 
galaxies in the local Universe populate the \vm--\sig\ plane 
in the manner observed.  In this sense, we can regard the distribution of 
galaxies on the \vm--\sig\ diagram as another local empirical boundary 
condition---much like other empirical galaxy scaling relations---that can 
be used to both guide and constrain theoretical models.  Courteau et al. 
(2007), for example, attempted to reproduce, with limited success, the actual 
variation of \vm/\sig\ with concentration index using the equilibrium 
disk-bulge-halo models of Widrow \& Dubinski (2005).

\subsection{Origin of the Low-\vm/\sig\ Systems}

One of the most notable features in the \vm--\sig\ diagram is the existence of 
a significant number of outliers with very low values of \vm/\sig\ ({\it open 
triangles}\ in Fig.~2).  As an operational definition, we have designated this 
population as those that deviate from the $K_s$-band Tully-Fisher relation 
toward the low-\vm\ side by more than twice the rms scatter.  This definition, 
though somewhat ad hoc, effectively isolates most of the members characterized 
by \vm/\sig\ \lax\ 1.  Although the low-\vm/\sig\ objects comprise only 17\% 
(132/792) of the total sample, they are nonetheless very intriguing because 
the kinematic measurements are robust.  Some of these systems have rotation 
velocities as low as $\sim$30 \kms, and yet most are luminous, with 
$M_{B_{\rm T}^c} \approx -19$ to $-21$ mag.  Moreover, these galaxies have 
central stellar velocity dispersions as large as $\sigma_0 \approx 100-250$ 
\kms: it is impossible for a cold disk rotating this slowly to survive in such 
a dynamically hot system.  The low-\vm/\sig\ outliers can be found in 
essentially all Hubble type bins included in our study, but they appear to be 
somewhat more prevalent among early-type (E and S0) galaxies: they make up 
33\% of the ellipticals, 40\% of the S0s, and $\sim 15$\% to 25\% of the 
spirals, depending on Hubble type.

Apart from their low \vm/\sig\ ratios, the outliers do not stand out in any 
other obvious manner, at least with the data we readily have at our disposal 
(using data given in Tables 1 and 2 and as listed in Hyperleda).  
When normalized with respect to other galaxies of the same morphological type, 
they appear to be normal in terms of absolute magnitude, mean surface 
brightness (measured either at the effective radius or within $\mu_B$ = 25 mag 
arcsec$^{-2}$), integrated broad-band colors ($U-B$, $B-V$, $B-I$, $B-K_s$), 
and \hi\ content (either the total \hi\ mass, \mhi, or the \hi\ mass normalized 
to the $B$-band luminosity, \mhi/\lb).  Among the disk galaxies in the sample, 
56\% are classified as barred, to be compared with 42\% for the kinematically
normal sample; it is unclear if the marginal excess of barred galaxies in the 
former sample is statistically significant.  Other than the indirect effect of 
the morphology-density relation (e.g., Dressler 1980), no obvious trends with 
environment exist, although except for noting cluster membership we did not 
attempt to quantify the environment in any rigorous manner.  

How might a luminous galaxy with a sizable hot spheroidal component have such 
kinematically cold neutral hydrogen?  The first and most obvious possibility 
is that we have severely misjudged the inclination correction to the \hi\ line 
width. Recall that the inclinations were derived from the axial ratios of the 
{\it stellar}\ isophotes, which implicitly assumes that the \hi\ and stars are 
coaligned.  If the gas and the stars are significantly misaligned, then the 
optically derived inclination angles may either overestimate or underestimate 
the inclination correction for the \hi\ line width, depending on the sense of 
the misalignment.  This then leads to inferred rotational velocities that 
are either too large or too small, with about equal probability.  This effect 
must occur at some level, as galaxies with \hi\ disks with varying degrees of 
misalignment are well-known, the most dramatic examples being polar ring 
systems, in which the gaseous and stellar disks are exactly orthogonal to one 
another.  That this effect does occur in our sample can be seen in the 
Tully-Fisher diagrams shown in Figure~3: most galaxies cluster around a 
central ridgeline, but a significant number of objects deviate both to the 
low-velocity and high-velocity side of the ridgeline.  A crucial point to 
note, however, is that the scatter is not symmetric: there is an {\it excess}\ 
of low-velocity sources compared to high-velocity sources.  Moreover, the 
absolute deviation from the Tully-Fisher ridgeline is more extreme for the 
low-velocity objects than the high-velocity objects.  Although \hi\ disk 
misalignment may account for some of the low-\vm/\sig\ objects, this is not 
the whole story.

The asymmetric distribution of the Tully-Fisher outliers compels us to 
conclude that the \hi\ in at least some of these objects must reside in a 
truly {\it dynamically unrelaxed}\ configuration with respect to the 
stellar gravitational potential.  Three possibilities come to mind.  First, 
the \hi\ gas may be strongly disturbed as a result tidal interactions, or 
perhaps distributed in tidal tails formed in the aftermath of a galaxy-galaxy 
merger.  From optical studies of close pairs and interacting galaxies, Barton 
et al. (2001) and Kannappan \& Barton (2004) discuss how asymmetric and 
truncated rotation curves can generate strong offsets in the Tully-Fisher 
relation.  While this effect might account for some fraction of the outliers, 
recall that during our sample selection we have purposefully excluded galaxies 
with close companions and obvious morphological peculiarities.  Thus, this
cannot offer a viable solution to the problem at hand.  

Second, perhaps the \hi\ owes its kinematic anomaly to a large-scale galactic 
outflow, driven either by strong nuclear activity or a global starburst.  We 
also consider this possibility to be untenable.  The few galaxies with 
luminous active galactic nuclei or starbursts in our study (e.g., NGC~1068, 
NGC~3079, NGC~4051)---the most promising candidates for this 
scenario---belong, in fact, to the kinematically normal sample, exactly the 
opposite of what is expected. Furthermore, although we do not have uniform 
data to estimate star formation rates for the entire sample, the similarity in 
the broad-band optical and near-infrared colors of the two groups strongly 
suggests that their star formation rates are not grossly dissimilar.  
Lastly, it is worth remarking that the amount of neutral gas involved is quite 
substantial.  The median \hi\ mass for the low-\vm/\sig\ sample ranges from 
$3\times10^8$ \solmass\ for the S0s to $\sim 1\times10^9$ \solmass\ for the 
Es, $2\times10^9$ \solmass\ for Sa spirals, and $\sim 1\times 10^{10}$ 
\solmass\ for spirals of type Sb and later.  Without detailed mapping we do 
not know precisely how much of the gas actually resides in the low-velocity 
component, but for it to markedly affect the integrated profile to the degree 
observed, the fraction of the total mass involved cannot be that miniscule.  It 
seems doubtful that nuclear outflows or galactic fountains can perturb such a 
large quantity of gas.  In NGC~891 (Swaters et al. 1997) and NGC~2403 
(Fraternali et al. 2002), two of the best-studied spirals whose kinematically 
anomalous \hi\ has been interpreted as possibly having a galactic fountain 
origin, the amount of gas affected is $\sim$few$\times 10^8$ \solmass, roughly 
10\% of the total \hi\ mass.  Moreover, this material rotates only a few tens 
of \kms\ slower than the gas in the plane, nowhere near as extreme as some of 
our objects.

The final and, we believe, most likely possibility is that the dynamically 
unrelaxed \hi\ traces material acquired from an external source, in the form 
of a minor merger with a gas-rich dwarf galaxy, capture of fallback material 
from incomplete conversion of gas to stars after a major merger, or cold 
accretion from primordial clouds.  Past studies of early-type galaxies have 
long invoked an external origin to account for the unusual properties of the 
neutral hydrogen in these systems, including the frequent detection of 
morphological and kinematic misalignments between the \hi\ and stars (e.g., 
Raimond et al. 1981; van~Gorkom et al. 1986; van~Driel \& van~Woerden 1991; 
Oosterloo et al. 2002, 2007; Noordermeer et al. 2005; Morganti et al. 2006) 
and the apparent decoupling between the \hi\ and stellar mass (Knapp et al. 
1985; Wardle \& Knapp 1986).  In this sense, our results for E and S0 galaxies 
could have been anticipated.  More surprising is that the same phenomenon 
persists well into the spiral sequence, to Hubble types as late as Scd.  The 
incidence of low-\vm/\sig\ objects does decrease toward later Hubble types, 
but this may be an artifact of dilution by the intrinsically much larger \hi\ 
content in these gas-rich systems.  If, for the sake of argument, external 
accretion provides a baseline supply of $10^8$ \solmass\ of \hi\ to all 
sizable galaxies, irrespective of Hubble type or environment, this 
kinematically anomalous component would constitute a significantly larger 
fraction of the total \hi\ budget in an E or S0 galaxy ($\sim 20\%-50\%$) than 
in an Sb or Sc galaxy ($\sim 1\%-5\%$), rendering it much more subtle to 
detect in the latter.

Notwithstanding the apparent similarity between early-type and spiral 
galaxies, the origin of the kinematically anomalous gas may be very different 
in these two classes of objects.  The characteristically denser environment 
around ellipticals and at least some S0 galaxies sets a natural stage for 
interaction or merger-induced processes to play a more central role than in 
later-type galaxies.  Spirals, on the other hand, may have access to another 
channel of gas supply---``cold accretion.''  In numerical simulations of galaxy 
formation, accretion of cold gas along filamentary structures dominates the 
growth of lower-mass galaxies at high redshifts and in low-density 
environments today (Kere\v{s} et al. 2005; Macci\'o et al. 2006).  Despite 
the purported cosmological importance of this process, there has been very 
limited observational evidence to support it.  A long-standing argument that 
spiral galaxies may accrete appreciable amounts of gas directly from the 
intergalactic medium comes from considerations of the high-velocity clouds in 
the Milky Way (Oort 1970; Wakker et al. 1999) and their probable extragalactic 
counterparts (van~der~Hulst \& Sancisi 1988; Thilker et al. 2004).  Using a 
blunt but effective tool, namely the \vm--\sig\ diagram in combination with 
the Tully-Fisher relation, this study has highlighted a population of galaxies 
with kinematically anomalous \hi\ that appear to be excellent candidates for 
undergoing cold accretion.  Deep aperture synthesis observations of {\it 
isolated}\ spirals are needed to confirm or reject this hypothesis; to our 
knowledge, none of the candidates from our list has been studied in this way.  
The selection by environment is important in order to distinguish a true cold 
accretion event from other sources of gas infall.  For example, the Sc spiral 
NGC~4254, one of the low-\vm/\sig\ galaxies in our sample, has long been known 
to possess kinematically distorted \hi, but its location in the Virgo cluster 
suggests that its peculiar gas kinematics and distribution most likely arose 
from a tidal encounter with another galaxy or from interactions with the 
intracluster medium (e.g., Phookun \& Mundy 1995; Vollmer et al. 2005).  

\vskip 0.5truein
\section{Summary}

We reinvestigate the relation between the maximum rotation velocity of the disk
and the central stellar velocity dispersion of the bulge in order to evaluate 
the claim that these two quantities are tightly correlated and independent of 
galaxy type.  Making use of integrated \hi\ line widths to estimate \vm\ and 
an expanded database of \sig\ values augmented by new measurements from SDSS, 
our sample contains 792 galaxies, almost a factor of 5 larger than any 
past study.  Contrary to previous reports, we find that the \vm--\sig\ 
relation contains significant intrinsic scatter, and that the ratio \vm/\sig\ 
varies systematically with Hubble type, bulge-to-disk ratio, and light 
concentration.  The density profile of a galaxy plays a major role in 
determining \vm/\sig, although the large residual scatter suggests that the 
kinematic structure of the galaxy also matters.    Extreme late-type spirals 
that lack a clear bulge but contain a central nuclear star cluster deviate 
dramatically from the \vm--\sig\ relation.  The observed distribution of 
galaxies on the \vm--\sig\ plane serves as an important local boundary 
condition to constrain models of galaxy formation.

The lack of a tight \vm--\sig\ relation removes the principal motivation for 
substituting the halo for the bulge as the galaxy component most closely 
linked with the central black hole.  Although \vm\ is inferior to \sig\ as a 
predictor of black hole mass, the \mbh--\vm\ relation remains a useful tool in 
instances when \sig\ is too difficult to measure.

To constrain the intrinsic distribution of \vm/\sig\ for any given Hubble 
type, we constructed a $K_s$-band Tully-Fisher relation for the sample using 
near-infrared photometry from 2MASS.  The $K_s$-band Tully-Fisher relation 
is essentially invariant, for morphological types ranging from elliptical 
galaxies to late-type spirals.  This exercise revealed an unexpected 
population of outliers characterized by having anomalously low rotation 
velocities for their luminosity, and correspondingly low \vm/\sig\ values.
While misaligned \hi\ disks and tidal tails may account for some of these
low-\vm/\sig\ objects, we argue that the majority of them must have acquired 
their low-velocity, dynamically unrelaxed gas through external capture or 
cold accretion from the intergalactic medium.

\acknowledgements
This work was supported by the Carnegie Institution of Washington and by NASA 
grants from the Space Telescope Science Institute (operated by AURA, Inc., 
under NASA contract NAS5-26555).  Extensive use was made of the databases in 
Hyperleda, the Sloan Digital Sky Survey, and the NASA/IPAC 
Extragalactic Database, which is operated by the Jet Propulsion 
Laboratory, California Institute of Technology, under contract with NASA.
I am especially grateful to David Schlegel for making publically available 
his database of SDSS stellar velocity dispersions, which added significantly 
to the data used in this paper.  I thank Jenny Greene for assistance with 
accessing the SDSS database.  I am grateful to St\'ephane Courteau, 
Jeremy Darling, and Jenny Greene for comments on an early draft of this paper.
I benefited from helpful correspondence with St\'ephane Courteau, Jeremy 
Darling, Sandra Faber, Jenny Greene, Daisuke Kawata, Gus Oemler, Karen O'Neil, 
Jacqueline van~Gorkom, and Bradley Whitmore.  An anonymous referee provided 
very helpful suggestions.

\clearpage
\appendix

\section{Comparison of Literature Data}

\subsection{Stellar Velocity Dispersions}

Among the 293 galaxies from our original sample 1 selected from Hyperleda (see 
\S 2), 37 have independent \sig\ measurements given in the  SDSS database.  
Figure~8{\it a}\ compares the velocity dispersion data in common.  The quoted 
error bars from SDSS are significantly smaller than those given in Hyperleda, 
but it is not clear how realistic the SDSS error values are.  Although the 
scatter is large, on average the SDSS velocity dispersions tend to be smaller 
than those given in Hyperleda; for this comparison sample, we find 
$\langle$\sig(SDSS)/\sig(Hyperleda)$\rangle=0.89\pm0.30$ (Fig.~8{\it b}).  
Because the Hyperleda velocity dispersion scale has been cross-calibrated 
using multiple measurements of bright galaxies (following the procedure of 
McElroy 1995), we believe that the Hyperleda values are more reliable.  
Consequently, we have increased the velocity dispersions for the
SDSS objects (sample 2) by a factor of 1.12.

\subsection{Rotation Velocities}

Reliable integrated \hi\ line widths are available for 93 of the galaxies for 
which Courteau et al. (2007) compiled rotation velocities measured from 
extended optical rotation curves.  A comparison of the two quantities is given 
in Figure~9{\it a}.  There is generally good agreement between the two sets of
measurements, but on average the \hi-based velocities are smaller by a factor 
of 1.09, with a standard deviation of 0.22.  Given that our values of \vm\ 
depend on a specific calibration between \hi\ line width and rotation velocity 
and a specific formalism for correcting the \hi\ line width for turbulent 
broadening (see \S 2), this level of agreement is quite satisfactory.  In 
any case, only a small fraction of our objects (4\%; total of 33 objects from 
samples 4b and 4c) come from Courteau et al.'s compilation.  Closer 
inspection reveals a number of prominent outliers, particularly toward low 
\hi\ velocities.  Figure~9{\it b}, which plots the ratio of \hi\ to optical 
velocities as a function of morphological type, illustrates that early-type 
($T$ \lax 0, or E and S0) galaxies show the greatest tendency for the \hi\ 
line widths to underpredict the optical rotation velocities.  The implications
of this trend are further discussed in \S 4.3.

We remarked in \S 2 that Hyperleda's procedure of homogenizing \hi\ line widths 
does not take into account the possibility of source confusion.  We illustrate 
this effect in Figure~10, where the Hyperleda velocities are compared directly 
with our rederived velocities.  Although most of the points agree reasonably 
well for \vm\ \gax 150--200 \kms, note that at lower velocities the Hyperleda 
values tend to be systematically and significantly (up to $\sim$40\%) larger 
than ours.  We attribute this effect to source confusion in the Hyperleda 
average.  Since the relative radial velocities of neighboring galaxies are 
generally larger than the internal velocity of any constituent galaxy, source 
confusion naturally leads to systematically larger line widths.

\subsection{Inclination Corrections}

In the course of tracking down the cause of highly discrepant values of 
\vm/\sig, we noticed that some of the inclination angles listed 
in Hyperleda are incorrect.  For instance, the nearby, well-resolved 
late-type spiral NGC~4395 is listed in Hyperleda as having $b/a = 0.37$ or 
$i = 90$\deg, whereas the Third Reference Catalogue of Bright Galaxies (RC3; 
de~Vaucouleurs et al. 1991) gives $b/a = 0.83$, which corresponds to 
$i = 38$\deg.  Simple inspection of the Digital Sky Survey images clearly 
shows that NGC~4395 is far from edge-on.  The S0 galaxy NGC~3106 provides 
another example: Hyperleda lists $b/a = 0.30$ and $i = 90$\deg, while the RC3 
gives $b/a = 1.0$, consistent with being nearly face-on, again in 
agreement with visual examination of the optical image of the galaxy.  While 
most galaxies in our sample show less blatant disagreement, we find that there 
is a general tendency for Hyperleda to underestimate the axial ratio ($b/a$), 
and hence to {\it overestimate}\ the inclination angle.  This is illustrated 
in Figure~11, where we use the RC3 as a reference.  Note the systematic 
difference between the two databases, and that the discrepancy is especially 
pronounced for barred galaxies.  While in principle there is no a priori 
reason to trust the RC3 more than Hyperleda, our systematic inspection of 
images for our sample (e.g., the two cases mentioned above) leads us to 
believe that the axial ratios given in the former are more reliable than those 
in the latter.

It is unclear why the axial ratios listed in Hyperleda are systematically 
lower.  In the case of barred galaxies, which show the largest systematic
deviations, it is conceivable that the high surface brightness of the bar may 
have biased the apparent ellipticity of the isophotes used to deduce the axial 
ratio.  In some instances, we noted that the final axial ratio adopted 
in Hyperleda does not, it seems, correspond to the $B$-band data actually 
tabulated in the database, from which the axial ratio (parameter {\tt logr25}) 
supposedly was derived.  Instead, it appears that the adopted axial ratio 
included in its average size measurements from near-infrared bandpasses.  In 
other cases (e.g., PGC~18506), the database does not, in fact, list any 
$B$-band measurements at all, even though {\tt logr25} strictly speaking is 
said to be derived from $B$-band data.  Since features such as 
bars tend to be more prominent in redder bandpasses, and shallow near-infrared 
surveys (e.g., DENIS; Paturel et al. 2005) are less effective at picking up 
the low-surface brightness outer regions of the disk component, barred 
galaxies will generally appear to have a higher ellipticity than if measured 
from deeper $B$-band photometry.  This probably contributes to the effect 
seen in Figure~11.

In light of these complications with the Hyperleda axial ratios, we decided to 
collect our own values from the literature (see Table 1).  Whenever possible, 
we give preference to the large body of uniform measurements given in the RC3.
We use these axial ratios to rederive inclination angles, following the 
same precepts adopted in Hyperleda (Paturel et al. 1997).  The inclination 
angles are then used to correct the \hi\ line widths, as well as to estimate 
internal extinctions for the absolute magnitudes, again following the 
procedures used in Hyperleda.

\vskip 0.3cm
\begin{figure*}[t]
\figurenum{8}
\centerline{\psfig{file=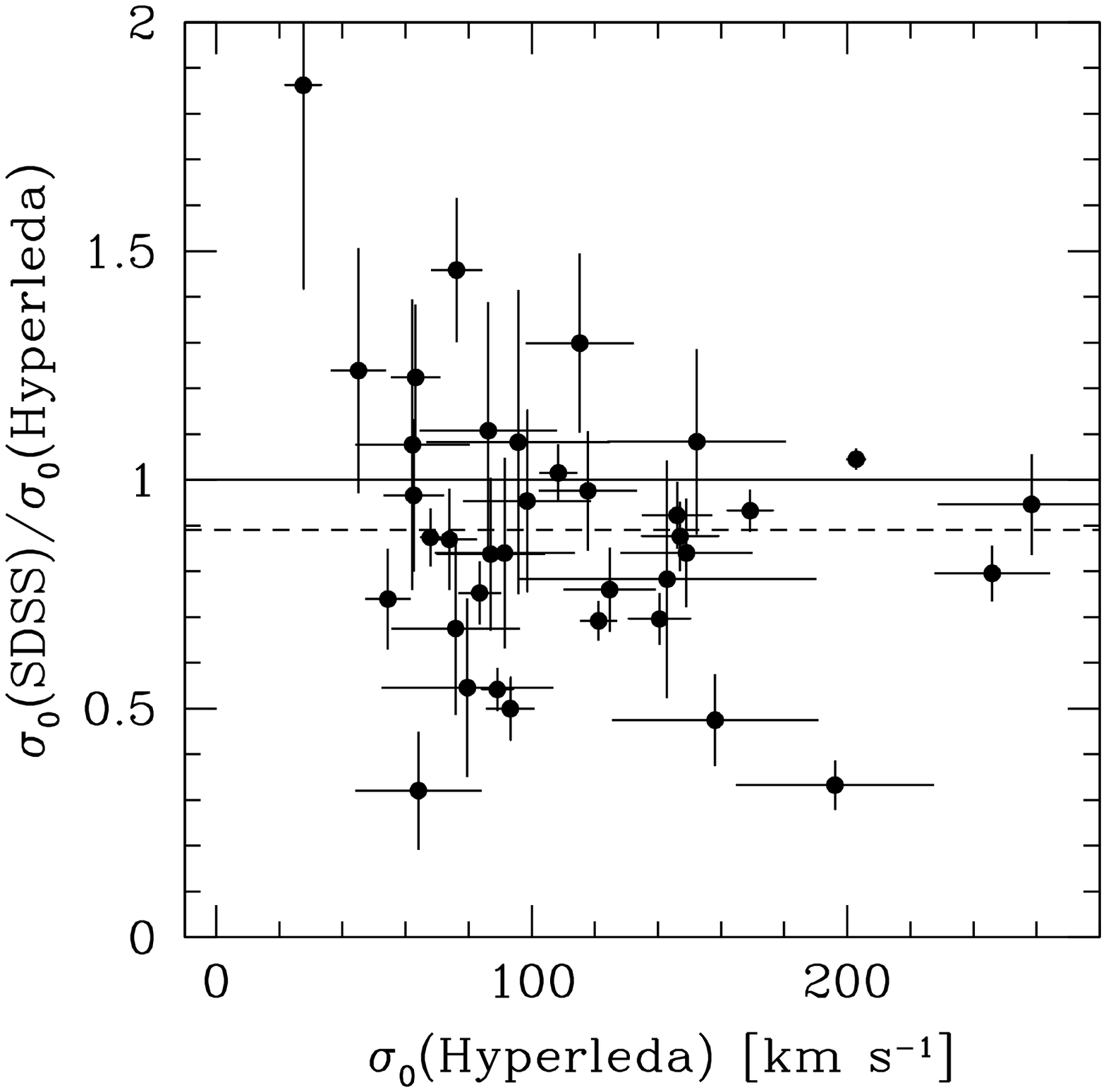,width=8.5cm,angle=0}}
\figcaption[fig8.ps]{Comparison of central velocity dispersions for 37
objects contained in both the Hyperleda sample (sample 1) and the SDSS sample
(sample 2).  On average
$\langle$\sig(SDSS)/\sig(Hyperleda)$\rangle=0.89\pm0.30$.
\label{fig8}}
\end{figure*}
\vskip 0.3cm

\vskip 0.3cm
\begin{figure*}[t]
\figurenum{9}
\centerline{\psfig{file=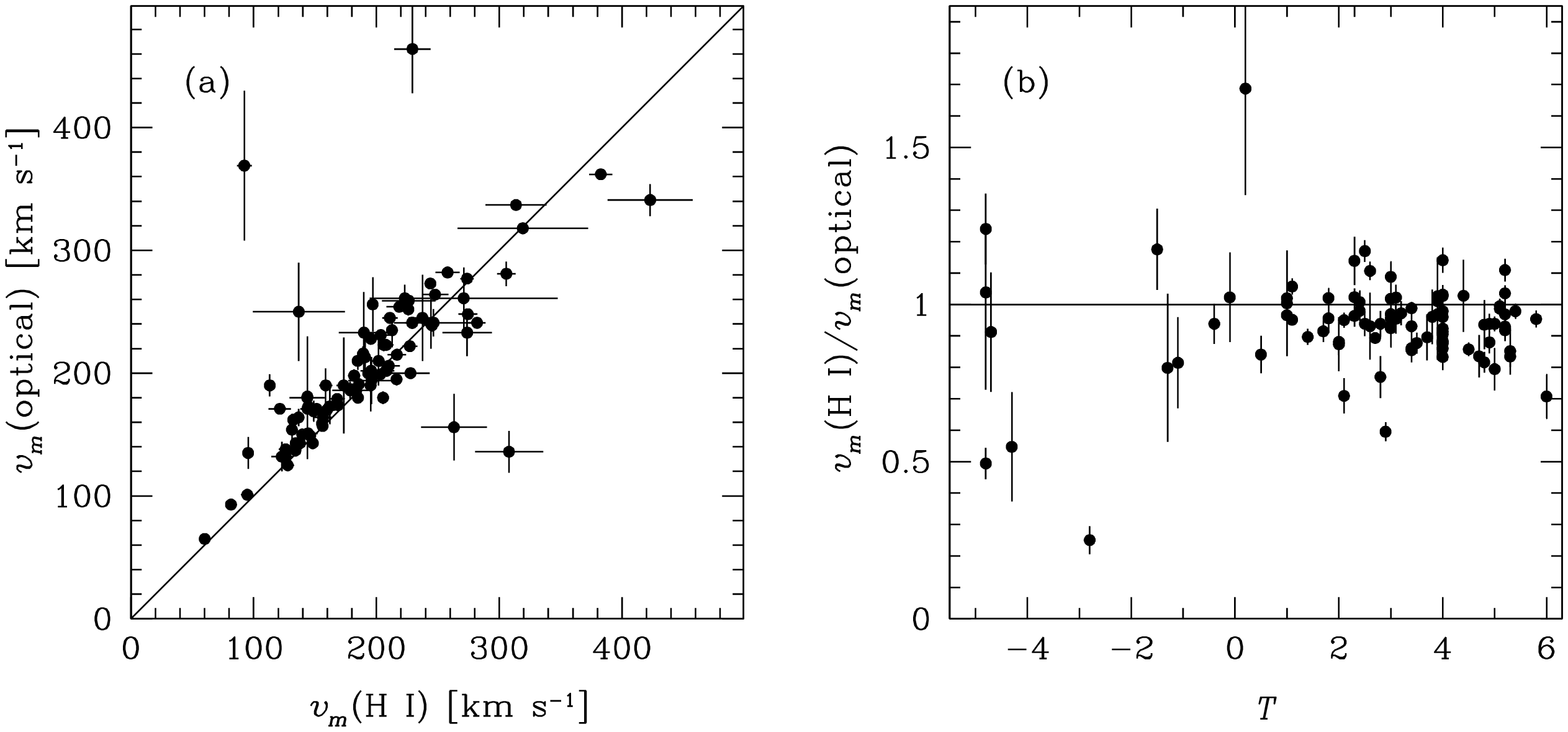,width=19.5cm,angle=-90}}
\figcaption[fig9.ps]{({\it a}) Comparison of rotational velocities obtained
from integrated \hi\ profiles with those measured from resolved optical
rotation curves, as compiled for 93 galaxies in the sample of Courteau et al.
(2007).  ({\it b}) Ratio of rotation velocities derived from \hi\ profile
versus resolved optical rotation curves as a function of morphological type.
Note the strong disagreement for early-type ($T$ \lax\ 0) systems.
\label{fig9}}
\end{figure*}
\vskip 0.3cm

\vskip 0.3cm
\begin{figure*}[t]
\figurenum{10}
\centerline{\psfig{file=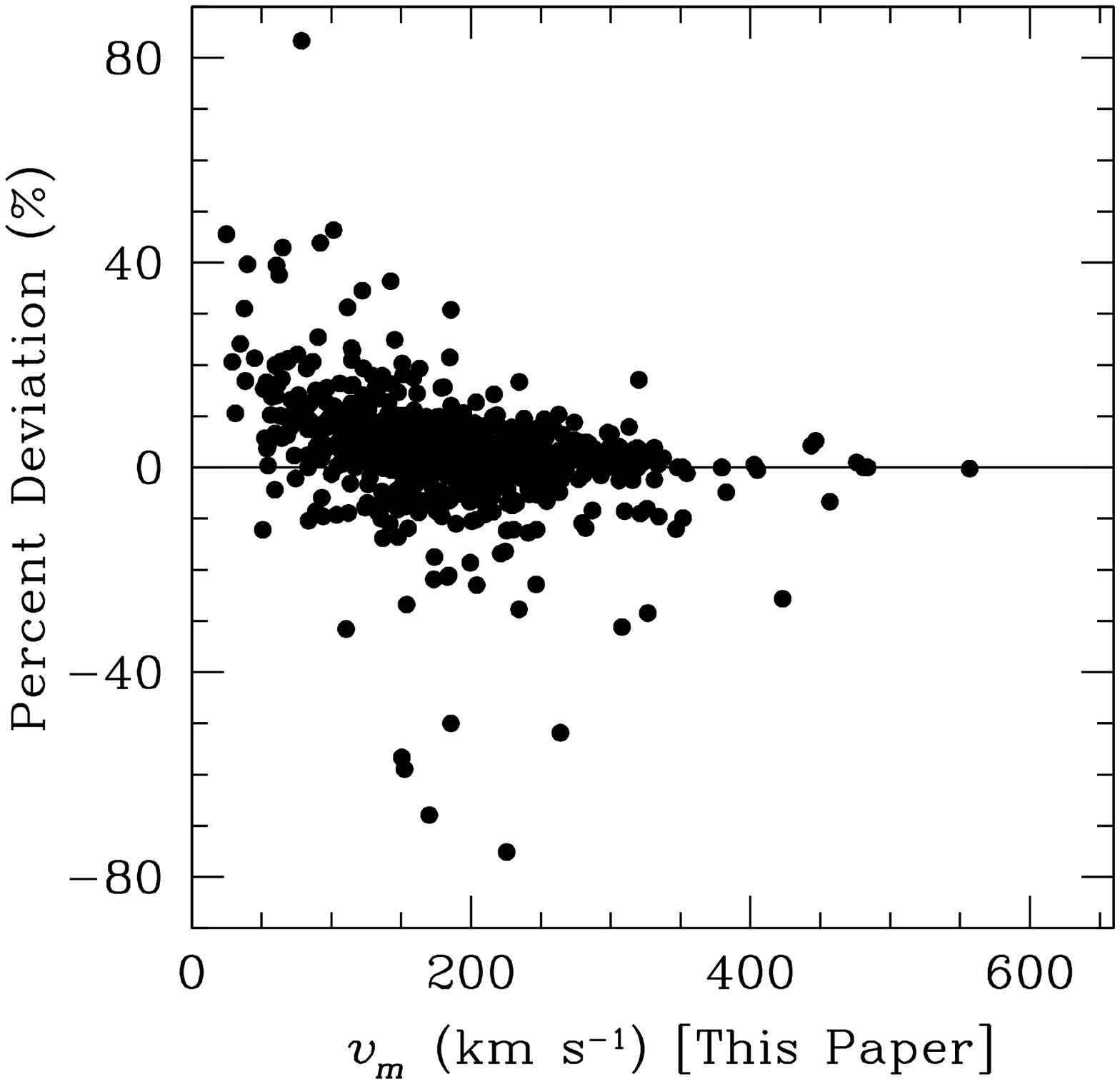,width=8.5cm,angle=0}}
\figcaption[fig10.ps]{Comparison of rotational velocities listed in Hyperleda
with those rederived in this paper using our own selection of \hi\ line widths.
The ordinate shows the percent deviation between the old and new values.
\label{fig10}}
\end{figure*}
\vskip 0.3cm

\vskip 0.3cm
\begin{figure*}[t]
\figurenum{11}
\centerline{\psfig{file=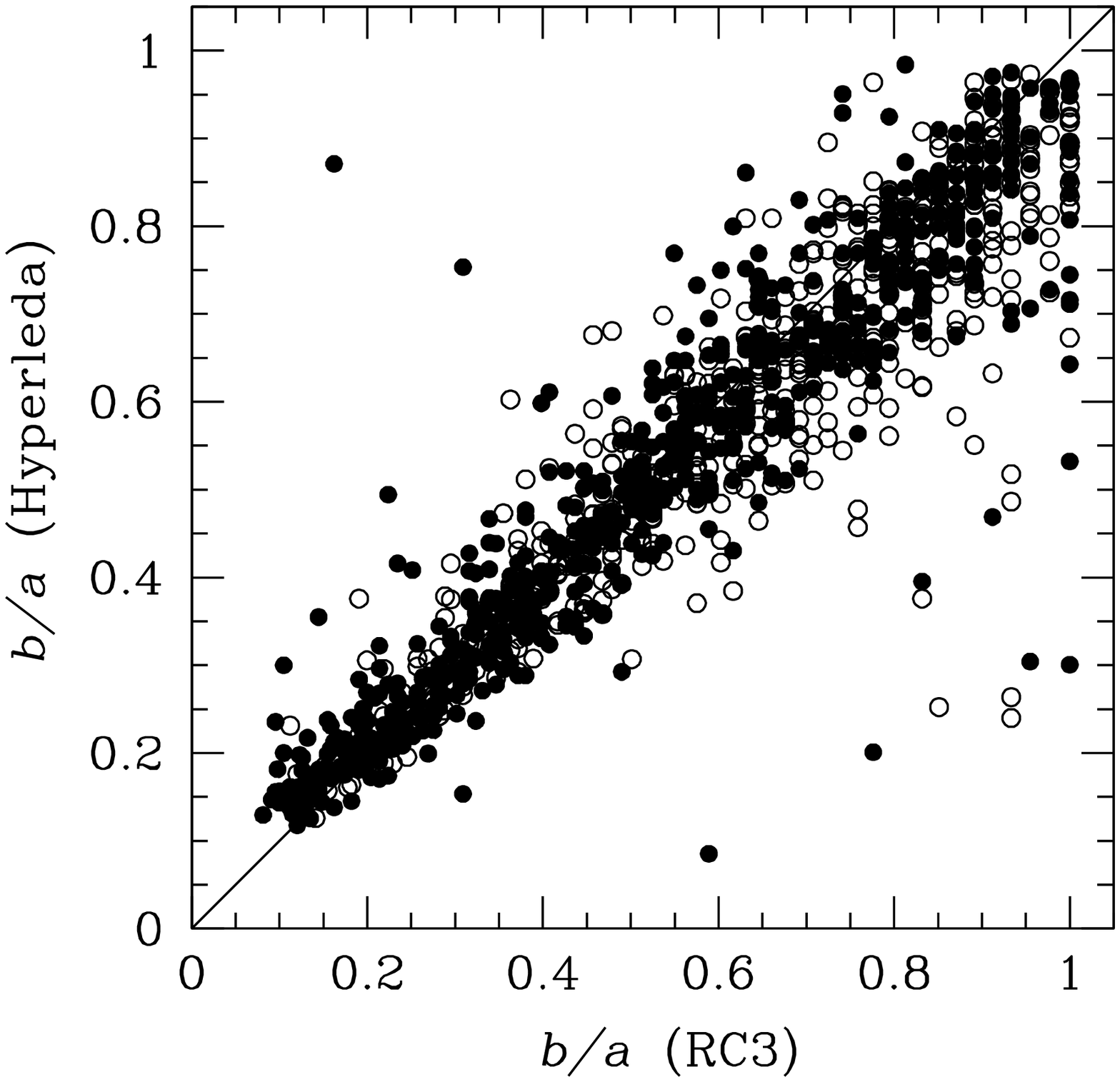,width=8.5cm,angle=0}}
\figcaption[fig11.ps]{Comparison of axial ratios ($b/a$) listed in Hyperleda
with those given in the RC3.  The ratio of semi-minor to semi-major
isophotal diameter is measured at a surface brightness of $\mu$ = 25 mag
arcsec$^{-2}$ in the $B$ band.  Unbarred and barred galaxies are denoted by
solid and open points, respectively.  The solid line marks a one-to-one
correspondence.
\label{fig11}}
\end{figure*}
\vskip 0.3cm


\begin{thebibliography}{}

\bibitem[]{1371}
Adelman-McCarthy, J. K., et al. 2007, \apjs, submitted

\bibitem[]{1374}
Baes, M., Buyle, P., Hau, G.~K.~T., \& Dejonghe, H. 2003, \mnras, 341, L44

\bibitem[]{1377}
Baes, M., Dejonghe, H., Buyle, P., Ferrarese, L., \& Gentile, G. 2004, in 
IAU Symp. 222, The Interplay among Black Holes, Stars and ISM in Galactic 
Nuclei, ed. T. Storchi-Bergmann, L. C. Ho, \& H. R. Schmitt (Cambridge: 
Cambridge Univ. Press), 25

\bibitem[]{1383}
Baldwin, J.~E., Lynden-Bell, D., \& Sancisi, R. 1980, \mnras, 193, 313

\bibitem[]{1386}
Balkowski, C., \& Chamaraux, P. 1981, \aa, 97, 223

\bibitem[]{1389}
------. 1983, \aas, 51, 331

\bibitem[]{1392}
Barnes, D. G., Staveley-Smith, L., Webster, R. L., \& Walsh, W. 1997, \mnras, 
288, 307

\bibitem[]{1396}
Barth, A.~J., Greene, J. E., \& Ho, L.~C. 2005, \apj, 619, L151

\bibitem[]{1399}
Barth, A.~J., Ho, L.~C., Rutledge, R. E., \& Sargent, W.~L.~W. 2004, \apj,
607, 90

\bibitem[]{1403}
Barton, E.~J., Geller, M.~J., Bromley, B.~C., van Zee, L., \& Kenyon, S.~J.  
2001, \aj, 121, 625

\bibitem[]{1407}
Bell, E.~F., \& de Jong, R.~S. 2001, \apj, 550, 212

\bibitem[]{1410}
Bernardi, M., et al. 2003, \aj, 125, 1849

\bibitem[]{1413}
Bessell, M.~S. 2005, \annrev, 43, 293

\bibitem[]{1416}
Bicay, M.~D., \& Giovanelli, R. 1986a, \aj, 91, 705

\bibitem[]{1419}
------. 1986b, aj, 91, 732

\bibitem[]{1422}
------. 1987, aj, 93, 1326

\bibitem[]{1425}
Bieging, J.~H., \& Biermann, P.  1977, \aa, 60, 361

\bibitem[]{1428}
Binggeli, B., Sandage, A.~R., \& Tammann, G.~A. 1985, \aj, 90, 1681

\bibitem[]{1431}
B\"oker, T., Stanek, R., \& van der Marel, R.~P. 2003, \aj, 125, 1073

\bibitem[]{1434}
B\"oker, T., van der Marel, R.~P., Laine, S., Rix, H.-W., Sarzi, M.,
Ho, L.~C., \& Shields, J.~C. 2002, \aj, 123, 1389

\bibitem[]{1438}
Bosma, A. 1979, Ph.D. Thesis, Univ. Groningen

\bibitem[]{1441}
------. 1981, \aj, 86, 1825

\bibitem[]{1444}
Bothun, G.~D., Aaronson, M., Schommer, B., Huchra, J., \& Mould, J. 1984,
\apj, 278, 475

\bibitem[]{1448}
Bothun, G.~D., Aaronson, M., Schommer, B., Mould, J., Huchra, J., \&
Sullivan, W. T., III 1985, \apjs, 57, 423

\bibitem[]{1452}
Bottinelli, L., \& Gouguenheim, L. 1977, \aa, 60, L23

\bibitem[]{1455}
------. 1979, \aa, 76, 176

\bibitem[]{1458}
Bottinelli, L., Gouguenheim, L., \& Paturel, G. 1980, \aa, 88, 32

\bibitem[]{1461}
------. 1982, \aa, 113, 61

\bibitem[]{1464}
Bottinelli, L., Gouguenheim, L., Paturel, G., \& de Vaucouleurs, G. 1983, 
\aa, 118, 4

\bibitem[]{1468}
Bottinelli, L., Gouguenheim, L., Paturel, G., \& Teerikorpi, P. 1995, \aa, 296, 64

\bibitem[]{1471}
Bottinelli, L., Gouguenheim, L., Theureau, G., Coudreau, N., \& Paturel, G. 
1999, \aas, 135, 429

\bibitem[]{1475}
Bregman, J.~N., Hogg, D.~E., \& Roberts, M.~S. 1992, \apj, 387, 484

\bibitem[]{1478}
Bregman, J.~N., \& Roberts, M. S. 1988, \apj, 330, L93

\bibitem[]{1481}
Broeils, A.~H., \& Rhee, M.-H. 1997, \aa, 324, 877

\bibitem[]{1484}
Burstein, D., Krumm, N., \& Salpeter, E.~E. 1987, \aj, 94, 883

\bibitem[]{1487}
Buyle, P., Ferrarese, L., Gentile, G., Dejonghe, H., Baes, M., \& Klein, U. 
2006, \mnras, 373, 700

\bibitem[]{1491}
Chamaraux, P., Balkowski, C., \& Fontanelli, P.  1987, \aas, 69, 263

\bibitem[]{1494}
Chamaraux, P., Cayatte, V., Balkowski, C., \& Fontanelli, P. 1990, \aa, 229, 340

\bibitem[]{1497}
Chengalur, J.~N., Salpeter, E.~E., \& Terzian, Y. 1994, \aj, 107, 1984

\bibitem[]{1500}
Courteau, S. 1997, \aj, 114, 2402

\bibitem[]{1503}
Courteau, S., McDonald, M., Widrow, L. M., \& Holtzman, J. 2007, \apj, 655, L21

\bibitem[]{1506}
Davies, R.~D., \& Lewis, B.~M. 1973, \mnras, 165, 231

\bibitem[]{1509}
Davis, L.~E., \& Seaquist E. R. 1983, \apjs, 53, 269

\bibitem[]{1512}
Dean, J. F., \& Davies, R. D. 1975, \mnras, 170, 503

\bibitem[]{1515}
Dell'Antonio, I., Bothun, G. D., \& Geller, M. J. 1996, \aj, 112, 1759

\bibitem[]{1518}
De Rijcke, S., Zeilinger, W.~W., Hau, G. K. T., Prugniel, P., \& Dejonghe, H. 
2007, \apj, 659, 1172

\bibitem[]{1522}
de Vaucouleurs, G., de Vaucouleurs, A., Corwin, H.~G., Jr., Buta, R.~J.,
Paturel, G., \& Fouqu\'e, R. 1991, Third Reference Catalogue of Bright
Galaxies (New York: Springer) (RC3)

\bibitem[]{1527}
Dickel, J.~R., \& Rood, H.~J. 1978, \apj, 223, 391

\bibitem[]{1530}
Doi, M., Fukugita, M., \& Okamura, S. 1993, \mnras, 264, 832

\bibitem[]{1533}
Dressler, A. 1980, \apj, 236, 351

\bibitem[]{1536}
Dutton, A. A., van den Bosch, F. C., Dekel, A., \& Courteau, S. 2007, 
\apj, 654, 27

\bibitem[]{1540}
Eder, J., Giovanelli, R., \& Haynes, M.~P. 1991, \aj, 102, 572

\bibitem[]{1543}
Faber, S.~M., \& Jackson, R.~E. 1976, \apj, 204, 668

\bibitem[]{1546}
Fall, S.~M. 1987, in Nearly Normal Galaxies, ed. S.~M. Faber (New York: 
Springer), 326

\bibitem[]{1550}
Ferrarese, L. 2002, \apj, 578, 90

\bibitem[]{1553}
Ferrarese, L., \& Merritt, D. 2000, \apj, 539, L9

\bibitem[]{1556}
Filippenko, A.~V., \& Ho, L.~C. 2003, \apj, 588, L13

\bibitem[]{1559}
Fisher, J. R., \& Tully, R. B. 1981, \apjs, 47, 139

\bibitem[]{1562}
Fontanelli, P. 1984, \aa, 138, 8

\bibitem[]{1565}
Fosbury, R.~A.~E., Mebold, U., Goss, W.~M., \& Dopita, M.~A. 1978, \mnras, 
183, 549

\bibitem[]{1569}
Fouqu\'e, R., Bottinelli, L., Gouguenheim, L., \& Paturel, G. 1990, \apj, 349, 1

\bibitem[]{1572}
Franx, M. 1993, in IAU Symp. 153, Galactic Bulges, ed. H. Dejonghe \& H.~J. 
Habing (Dordrecht: Kluwer), 243

\bibitem[]{1576}
Fraternali, F., van Moorsel, G., Sancisi, R., \& Oosterloo, T. 2002, \aj, 
123, 3124

\bibitem[]{1580}
Freudling, W. 1995, \aas, 112, 429

\bibitem[]{1583}
Freudling, W., Haynes, M. P., \& Giovanelli, R. 1988, AJ, 96, 1791

\bibitem[]{1586}
------. 1992, \apjs, 79, 157

\bibitem[]{1589}
Fukugita, M., Shimasaku, K., \& Ichikawa, T. 1995, \pasp, 107, 945

\bibitem[]{1592}
Garcia, A.~M., Bottinelli, L., Garnier, R., Gouguenheim, L., \& Paturel, 
G. 1994, \aas, 107, 265

\bibitem[]{1596}
Garc\'\i a-Barreto, J.~A., Downes, D., \& Huchtmeier, W.~K. 1994, \aa, 288, 705

\bibitem[]{1599}
Gavazzi, G. 1987, \apj, 320, 96

\bibitem[]{1602}
Gebhardt, K., et al.  2000, \apj, 539, L13

\bibitem[]{1605}
------. 2001, \aj, 122, 2469

\bibitem[]{1608}
Gerhard, O.~E., Kronawitter, A., Saglia, R.~P., \& Bender, R. 2001, \aj, 121, 
1936

\bibitem[]{1612}
Giovanardi, C., Krumm, N., \& Salpeter, E. E. 1983, AJ, 88, 1719

\bibitem[]{1615}
Giovanardi, C., \& Salpeter, E. E. 1985, \apjs, 58, 623

\bibitem[]{1618}
Giovanelli, R.,  Avera, E., \& Karachentsev, I. D. 1997, \aj, 114, 122

\bibitem[]{1621}
Giovanelli, R., Chincarini, L., \& Haynes, M.~P. 1981, \apj, 247, 383

\bibitem[]{1624}
Giovanelli, R., \& Haynes, M.~P. 1983, \aj, 88, 881

\bibitem[]{1627}
------. 1985a, \apj, 292, 404

\bibitem[]{1630}
------. 1985b, \aj, 90, 2445

\bibitem[]{1633}
------. 1993, \aj, 105, 1271

\bibitem[]{1636}
Greene, J. E., \& Ho, L. C. 2004, ApJ, 610, 722

\bibitem[]{1639}
------. 2006, ApJ, 641, 117

\bibitem[]{1642}
------. 2007a, ApJ, in press (astro-ph/0705.0020)

\bibitem[]{1645}
------. 2007b, submitted

\bibitem[]{1648}
Haynes, M.~P. 1981, \aj, 86, 1126

\bibitem[]{1651}
Haynes, M.~P., \& Giovanelli, R. 1984, \aj, 89, 758

\bibitem[]{1654}
------. 1991a, AJ, 102, 841

\bibitem[]{1657}
------. 1991b, \apjs, 77, 331

\bibitem[]{1660}
Haynes, M.~P., Giovanelli, R., Chamaraux, P., da Costa, L.~N., Freudling,
W., Salzer, J.~J., \& Wegner, G. 1999, \aj, 117, 2039

\bibitem[]{1664}
Haynes, M.~P., Giovanelli, R., Herter, T., Vogt, N. P., Freudling, W.,
Maia, M. A. G., Salzer, J. J., \& Wegner, G. 1997, AJ, 113, 1197

\bibitem[]{1668}
Haynes, M. P., Giovanelli, R., Starosta, B. M., \& Magri, C. 1988, 
\aj, 95, 606

\bibitem[]{1672}
Haynes, M.~P., Herter, T., Barton, A.~S., \& Benensohn, J.~S. 1990, 
\aj, 99, 1740

\bibitem[]{1676}
Haynes, M.~P., van Zee, L., Hogg, D. E., Roberts, M. S., \& Maddalena, R. J. 
1998, AJ, 115, 62

\bibitem[]{1680}
Heckman, T.~M., \& Balick, B., \& Sullivan, W.~T. 1978, \apj, 224, 745

\bibitem[]{1683}
Heckman, T.~M., Kauffmann, G., Brinchmann, J., Charlot, S., Tremonti, C.,
\& White, S. D. M. 2004, \apj, 613, 109

\bibitem[]{1687}
Heiles, C., et al. 2000, Arecibo Technical and Operations Memo 2000-04

\bibitem[]{1690}
Helou, G., Giovanardi, C., Salpeter, E. E., \& Krumm, N. 1981, \apjs, 46, 267

\bibitem[]{1693}
Helou, G., Hoffman, G. L., \& Salpeter, E. E. 1984, \apjs, 55, 433

\bibitem[]{1696}
Hewitt, J. N., Haynes, M. P., \& Giovanelli, R. 1983, \aj, 88, 272

\bibitem[]{1699}
Ho, L.~C. 2004, in Carnegie Observatories Astrophysics Series, Vol. 1:
Coevolution of Black Holes and Galaxies, ed. L. C. Ho (Cambridge: Cambridge
Univ. Press), 292

\bibitem[]{1704}
Ho, L. C., Darling, J., \& Greene, J. E. 2007a, \apjs, submitted

\bibitem[]{}
------. 2007a, \apj, submitted

\bibitem[]{1707}
Ho, L.~C., Filippenko, A.~V., \& Sargent, W.~L.~W. 1997, \apj, 487, 568

\bibitem[]{1710}
Hoffman, G.~L., Helou, G., \& Salpeter, E.~E. 1984, \apjs, 55, 433

\bibitem[]{1713}
Hoffman, G.~L., Lewis, B. M., Helou, G., Salpeter, E. E., \& Williams, H. L. 
1989, \apjs, 69, 55

\bibitem[]{1717}
Hoffman, G.~L., Lewis, B. M., Salpeter, E. E. 1995, \apj, 441, 28

\bibitem[]{1720}
Hopp, U., Kuhn, B., Thiele, U., Birkle, K., Elsasser, H., \& Kovachev, B. 
1995, \aas, 109, 537

\bibitem[]{1724}
Hubble, E. 1926, \apj, 64, 321

\bibitem[]{1727}
Huchtmeier, W.~K. 1973, \aa, 22, 91

\bibitem[]{1730}
------. 1982, \aa, 110, 121

\bibitem[]{1733}
Huchtmeier, W.~K., Hopp, U., \& Kuhn, B. 1997, \aa, 319, 67

\bibitem[]{1736}
Huchtmeier, W.~K., \& Richter, O.~G. 1985, \aa, 149, 118

\bibitem[]{1739}
------. 1987, \aas, 63, 323

\bibitem[]{1742}
Huchtmeier, W.~K., Sage, L.~J., \& Henkel, C. 1995, \aa, 300, 675

\bibitem[]{1745}
Huchtmeier, W.~K., \& Seiradakis, J. H. 1985, \aa, 143, 216

\bibitem[]{1748}
Impey, C., Burkholder, V., \& Sprayberry, D. 2001, \aj, 122, 2341

\bibitem[]{1751}
Irwin, J.~A., \& Seaquist, E. R. 1991, \apj, 371, 111

\bibitem[]{1754}
J{\o}rgensen, I., Franx, M., \& Kj\ae rgaard, P. 1995, \mnras, 276, 1341

\bibitem[]{1757}
Kannappan, S.~J., \& Barton, E. J. 2004, \aj, 127, 2694

\bibitem[]{1760}
Karachentsev, I.~D., et al. 2003, \aa, 404, 93

\bibitem[]{1763}
Kere\v{s}, D., Katz, N., Weinberg, D. H., \& Dav\'e, R. 2005, \mnras, 363, 2

\bibitem[]{1766}
Knapp, G.~R., Faber, S.~M., \& Gallagher, J.~S. 1978, \aj, 83, 11

\bibitem[]{1769}
Knapp, G.~R., Kerr, F.~J., \& Henderson, A. P. 1979, \apj, 234, 448

\bibitem[]{1772}
Knapp, G.~R., Turner, E.~L., \& Cunniffe, P.~E. 1985, \aj, 90, 454

\bibitem[]{1775}
Knapp, G.~R., van Driel, W., \& van Woerden, H. 1985, \aa, 142, 1

\bibitem[]{1778}
Koribalski, B. S., et al. 2004, \aj, 128, 16

\bibitem[]{1781}
Kormendy, J. 2004, in Carnegie Observatories Astrophysics Series, Vol. 1:
Coevolution of Black Holes and Galaxies, ed. L. C. Ho (Cambridge: Cambridge
Univ. Press), 1

\bibitem[]{1786}
Kormendy, J., \& Illingworth, G. 1983, \apj, 265, 632

\bibitem[]{1789}
Kormendy, J., \& Gebhardt, K. 2001, in The 20th Texas Symposium on Relativistic Astrophysics, ed. H. Martel \& J.~C. Wheeler (Melville: AIP), 363

\bibitem[]{1792}
Kormendy, J., \& Kennicutt, R. C.  2004, \annrev, 42, 603

\bibitem[]{1795}
Kormendy, J., \& McClure, R.~D. 1993, \aj, 105, 1793

\bibitem[]{1798}
Kraan-Korteweg, R.~C., van Driel, W., Briggs, F., Binggeli, B., \& Mostefaoui, 
T. I. 1999, \aas, 135, 255

\bibitem[]{1802}
Kronawitter, A., Saglia, R.~P., Gerhard, O., \& Bender, R. 2000, \aas, 144, 53

\bibitem[]{1805}
Krumm, N., \& Salpeter, E.~E. 1976, \apj, 208, L7

\bibitem[]{1808}
------. 1980, \aj, 85, 1312

\bibitem[]{1811}
Lee, M.~G., Kim, M., Sarajedini, A., Geisler, D., \& Wolfgang, G. 2002, \apj, 
565, 959

\bibitem[]{1815}
Lewis, B.~M. 1983, \aj, 88, 1695

\bibitem[]{1818}
------. 1987, \apjs, 63, 515

\bibitem[]{1821}
Lewis, B.~M., \& Davies, R. D. 1973, \mnras, 165, 213

\bibitem[]{1824}
Lewis, B.~M., Helou, G., \& Salpeter, E. E. 1985, \apjs, 59, 151

\bibitem[]{1827}
Lu, N.~Y., Dow, M. W., Houck, J. R., Salpeter, E. E., \& Lewis, B. M. 
1990, \apj, 357, 388

\bibitem[]{1831}
Lu, N.~Y., Hoffman, G.~L., Groff, T., Roos, T., \& Lamphier, C. 1993, \apjs, 
88, 383

\bibitem[]{1835}
Macci\'o, A. V., Moore, B., \& Stadel, J. 2006, \apj, 636, L25

\bibitem[]{1838}
MacGillivray, H. T., Beard, S. M., \& Dodd, R. J. 1988, in Astronomy from 
Large Databases: Scientific Objectives and Methodological Approaches 
(Garching: European Southern Observatory), 389

\bibitem[]{1843}
Magorrian, J., et al.  1998, \aj, 115, 2285

\bibitem[]{1846}
Magri, C. 1994, \aj, 108, 896

\bibitem[]{1849}
Marconi, A., \& Hunt, L.~K. 2003, \apj, 589, L21

\bibitem[]{1852}
Martin, J. M., Bottinelli, L., Gouguenheim, L., \& Dennefeld, M. 1991, 
\aa, 245, 393

\bibitem[]{1856}
Mathewson, D.~S., \& Ford, V.~L. 1996, \apjs, 107, 97

\bibitem[]{1859}
Mathewson, D.~S., Ford, V.~L., \& Buchhorn, M. 1992, \apjs, 81, 413

\bibitem[]{1862}
Matthews, L.~D., \& van Driel, W. 2000, \aas, 143, 421

\bibitem[]{1865}
Matthews, L.~D., van Driel, W., \& Monnier-Ragaigne, D. 2001, \aa, 365, 1

\bibitem[]{1868}
McElroy, D.~B. 1995, \apjs, 100, 105

\bibitem[]{1871}
Merritt, D., Ferrarese, L., \& Joseph, C.~L. 2001, Science, 293, 1116

\bibitem[]{1874}
Mirabel, I.~F., \& Sanders, D.~B. 1988, \apj, 335, 104

\bibitem[]{1877}
Mirabel, I.~F., \& Wilson, A.~S. 1984, \apj, 277, 92

\bibitem[]{1880}
Morganti, R., et al. 2006, \mnras, 371, 157

\bibitem[]{1883}
Mould, J.~R., et al. 1991, \apj, 383, 467

\bibitem[]{1886}
Mould, J. R., Akeson, R. L., Bothun, G. D., Han, M., Huchra, J. P., 
Roth, J., \& Schommer, R. A. 1993, \apj, 409, 14

\bibitem[]{1890}
Navarro, J.~F., \& Steinmetz, M. 2000, \apj, 538, 477

\bibitem[]{1893}
Nelson, C.~H., \& Whittle, M. 1996, \apj, 465, 96

\bibitem[]{1896}
Nilson, P. 1973, Upsala General Catalogue of Galaxies
(Uppsala: Astronomiska Observatorium)

\bibitem[]{1900}
Noordermeer, E., van der Hulst, J. M., Sancisi, R., Swaters, R. A., \& 
van Albada, T. S. 2005, \aa, 442, 137:

\bibitem[]{1904}
Nordgren, T.~E., Chengalur, J.~N., Salpeter, E.~E., \& Terzian, Y. 1998, 
\apjs, 115, 43

\bibitem[]{1908}
Oort, J.~H. 1970, \aa, 7, 381

\bibitem[]{1911}
Oosterloo, T.~A., Morganti, R., Sadler, E., Vergani, D., \& Caldwell, N.  
2002, \aj, 123, 729

\bibitem[]{1915}
Oosterloo, T.~A., Morganti, R., Sadler, E. M., van der Hulst, J. M., \& Serra, 
P. 2007, \aa, 465, 787

\bibitem[]{1919}
Oosterloo, T.~A., \& Shostak, S. 1993, \aas, 99, 379

\bibitem[]{1922}
Paturel, G., et al.  1997, \aas, 124, 109

\bibitem[]{1925}
Paturel, G., Fang, Y., Petit, C., Garnier, R., \& Rousseau, J. 2000, \aas, 146, 19

\bibitem[]{1928}
Paturel, G., Petit, C., Prugniel, Ph., Theureau, G., Rousseau, J., Brouty, M., 
Dubois, P., \& Cambr\'esy, L. 2003a, \aa, 412, 45

\bibitem[]{1932}
Paturel, G., Theureau, G., Bottinelli, L., Gouguenheim, L., Coudreau-Durand, 
N., Hallet, N.,  \& Petit, C. 2003b, \aa, 412, 57

\bibitem[]{1936}
Paturel, G., Vauglin, I., Petit, C., Borsenberger, J., Epchtein, N., 
Fouqu\'e, P., \& Mamon, G. 2005, \aa, 430, 751

\bibitem[]{1940}
Peterson, S.~D. 1979, \apjs, 40, 527

\bibitem[]{}
Phookun, B., \& Mundy, L. G. 1995, \apj, 453, 154

\bibitem[]{1943}
Pizagno, J., et al. 2007, \aj, in press (astro-ph/0608472)

\bibitem[]{1946}
Pizzella, A., Corsini, E., Dalla Bont\'a, E., Sarzi, M., Coccato, L., \& 
Bertola, F. 2005, \apj, 631, 785

\bibitem[]{1950}
Raimond, E., Faber, S.~M., Gallagher, J.~S., III, \& Knapp, G.~R.  1981, 
\apj, 246, 708

\bibitem[]{1954}
Reif, K., Mebold, U., Goss, W.~M., van Woerden, H., \& Siegman, B. 1982, 
\aas, 50, 451

\bibitem[]{1958}
Richter, O.-G., \& Huchtmeier, W.~K. 1982, \aa, 109, 155
 
\bibitem[]{1961}
------. 1987, \aas, 68, 427

\bibitem[]{1964}
------. 1991, \aas, 87, 425

\bibitem[]{1967}
Richter, O.-G., \& Sancisi, R. 1994, \aa, 290, L9

\bibitem[]{1970}
Rizzi, L., Bresolin, F., Kudritzki, R.-P., Gieren, W., Pietrzyski, G. 2006, 
\apj, 638, 766

\bibitem[]{1974}
Roberts, M.~S. 1978, \aj, 83, 1026

\bibitem[]{1977}
Roberts, M.~S., \& Haynes, M.~P. 1994, \annrev, 32, 115

\bibitem[]{1980}
Rosenberg, J.~L., \& Schneider, S.~E. 2000, \apjs, 130, 177

\bibitem[]{1983}
Roth, J., Mould, J. R., \& Davies, R. D. 1991, \aj, 102, 1303

\bibitem[]{1986}
Roth, J., Mould, J. R., \& Staveley-Smith, L. 1994, \aj, 108, 851

\bibitem[]{1989}
Rubin, V.~C., Burstein, D., Ford, W.~K., Jr., \& Thonnard, N. 1985, \apj, 
289, 81

\bibitem[]{1993}
Rubin, V.~C., Ford, W.~K., Jr., \& Thonnard, N. 1978, \apj, 225, L107

\bibitem[]{1996}
Rubin, V.~C., Ford, W.~K., Jr., Thonnard, N., Roberts, M. S., \& Graham,
J. A. 1976, \aj, 81, 687

\bibitem[]{2000}
Salzer, J.~J. 1992, \aj, 103, 385

\bibitem[]{2003}
Schneider, S.~E., Helou, G., Salpeter, E.~E., \& Terzian, Y. 1986, \aj, 
92, 742

\bibitem[]{2007}
Schneider, S.~E., Thuan, T. X., Magri, C., \& Wadiak, J. E. 1990, \apjs, 72, 245

\bibitem[]{2010}
Schneider, S.~E., Thuan, T. X., Mangum, J. G., \& Miller, J. 1992, \apjs, 81, 5

\bibitem[]{2013}
Schombert, J.~M. Bothun, G.~D., Schneider, S.~E., \& McGaugh, S.~S.  1992, 
\aj, 103, 1107

\bibitem[]{2017}
Shimasaku, K., et al. 2001, \aj, 122, 1238

\bibitem[]{2020}
Skrutskie, M. F., et al. 2006, \aj, 131, 1163

\bibitem[]{2023}
Sprayberry, D., Impey, C.~D., Irwin, M.~J., \& Bothun, G.~D. 1997, \apj, 
482, 104

\bibitem[]{2027}
Staveley-Smith, L., \& Davies, R. D. 1987, \mnras, 224, 953

\bibitem[]{2030}
------. 1988, \mnras, 231, 833

\bibitem[]{2033}
Staveley-Smith, L., Davies, R. D., \& Kinman, T. D. 1992, \mnras, 258, 334

\bibitem[]{2036}
Strateva, I., et al. 2001, \aj, 122, 1861

\bibitem[]{2039}
Sulentic, J.~W., \& Arp, H. 1983, \aj, 88, 489

\bibitem[]{2042}
Swaters, R.~A., Sancisi, R., \& van der Hulst, J. M. 1997, \apj, 491, 140

\bibitem[]{2045}
Theureau, G., et al. 2005, \aa, 430, 373

\bibitem[]{2048}
Theureau, G., Bottinelli, L., Coudreau-Durand, N., Gouguenheim, L., Hallet, N., Loulergue, M., Paturel G., \& Teerikorpi, P. 1998, \aas, 130, 333

\bibitem[]{2051}
Thilker, D.~A., Braun, R., Walterbos, R.~A.~M., Corbelli, E., Lockman, F. J., 
Murphy, E., \& Maddalena, R. 2004, \apj, 601, L39

\bibitem[]{2055}
Thim, F., Hoessel, J. G., Saha, A., Claver, J., Dolphin, A., \& Tammann,
G. A. 2004, \aj, 127, 2322

\bibitem[]{2059}
Thonnard, N. 1983, in Internal Linematics and Dynamics of Galaxies, ed. E. 
Athanassoula (Dordrecht: Reidel), 29

\bibitem[]{2063}
Thuan T.~X., Lipovetsky V.~A., Martin J.~M., \& Pustilnik S.~A. 1999, \aas,
139, 1

\bibitem[]{2067}
Thuan, T.~X., \& Martin, G. E. 1981, \apj, 247, 823

\bibitem[]{2070}
Tifft, W.~G., \& Cocke, W. J. 1988, \apjs, 67, 1

\bibitem[]{2073}
Tremaine, S., et al. 2002, \apj, 574, 740

\bibitem[]{2076}
Tully, R.~B., \& Fisher, J.~R. 1977, \aa, 54, 661

\bibitem[]{2079}
Tully, R.~R., \& Fouqu\'e, P. 1985, \apjs, 58, 67

\bibitem[]{2082}
Ulvestad, J.~S., \& Ho, L.~C. 2002, \apj, 581, 925

\bibitem[]{2085}
Valluri, M., Ferrarese, L., Merritt, D., \& Joseph, C. L. 2005, \apj, 628,
137

\bibitem[]{2089}
van der Hulst, T., \& Sancisi, R. 1988, \aj, 95, 1354

\bibitem[]{2092}
van Driel, W., Arnaboldi, M., Combes, F., \& Sparke, L.~S. 2000, \aas, 141, 385

\bibitem[]{2095}
van Driel, W., Marcum, P., Gallagher, J.~S., III, Wilcots, E., Guidoux,
 C., \& Ragaigne, D.~M.  2001, \aa, 378, 370

\bibitem[]{2099}
van Driel, W., Ragaigne, D., Boselli, A., Donas, J., \& Gavazzi, G. 2000,
 \aas, 144, 463

\bibitem[]{2103}
van Driel, W., \&  van Woerden, H. 1991, \aa, 243, 7

\bibitem[]{2106}
van Gorkom, J.~H., Knapp, G.~R., Raimond, E., Faber, S.~M., \& Gallagher, J.~S. 1986, \aj, 91, 791

\bibitem[]{2109}
Verheijen, M.~A.~W. 2001, \apj, 563, 694

\bibitem[]{2112}
Verheijen, M.~A.~W., \& Sancisi, R. 2001, \aa, 370, 765

\bibitem[]{}
Vollmer, B., Huchtmeier, W., \& van Driel, W. 2005, \aa, 439, 921

\bibitem[]{2115}
Vorontsov-Velyaminov, B. A., Arkipova, V. P., \& Kranogorskaja, A. A. 
1963-1974, Morphological Catalogue of Galaxies, Part I--V (Moscow: Trudy 
Sternberg Stat. Astr. Inst.)

\bibitem[]{2120}
Walcher, C.~J., et al. 2005, \apj, 618, 237 (err: 618, 237)

\bibitem[]{2123}
Wakker, B.~P., et al. 1999, \nat, 402, 388

\bibitem[]{2126}
Wardle, M., \& Knapp, G.~R. 1986, \aj, 91, 23

\bibitem[]{2129}
Wegner, G., Haynes, M. P., \& Giovanelli, R. 1993, \aj, 105, 1251

\bibitem[]{2132}
Whitmore, B.~C., \& Kirshner, R.~P. 1981, \apj, 250, 43A

\bibitem[]{2135}
Whitmore, B.~C., Schechter, P. L., \& Kirshner, R. P. 1979, \apj, 234, 68

\bibitem[]{2138}
Whittle, M. 1992a, \apj, 387, 109

\bibitem[]{2141}
------. 1992b, \apj, 387, 121

\bibitem[]{2144}
Widrow, L.~M. \& Dubinski, J. 2005, \apj, 631, 838

\bibitem[]{2147}
Williams, B.~A. 1985, \apj, 290, 462

\bibitem[]{2150}
Williams, B.~A., \& Kerr, F. J. 1981, \aj, 86, 953

\bibitem[]{2153}
Williams, B.~A., \& Rood, H. J. 1987, \apjs, 63, 265

\bibitem[]{2156}
York, D.~G., et al. 2000, \aj, 120, 1579

\bibitem[]{2159}
Young, C.~K., \& Currie, M.~J. 1998, \aas, 127, 367

\bibitem[]{2162}
Zasov, A. V.,  Petrochenko, L. N., \& Cherepashchuk, A. M. 2005, ARep, 49, 362 

\end{thebibliography}
\end{document}